%% file: paper.tex
\documentclass{art}

\usepackage{environ, multirow, tabularx, caption, booktabs, makecell, colortbl, joinargs}
\usepackage[frozencache=true,cachedir=.]{minted}
\setminted[]{fontfamily=lmtt}
\usepackage[prefix=bonak]{xkeymask}

\usetikzlibrary{patterns}

\makeatletter
\define@cmdkey[bonak]{X}{D}{(#1)}

\define@cmdkey[bonak]{frame}{D}{(#1)}
\define@cmdkey[bonak]{layer}{D}{(#1)}
\define@cmdkey[bonak]{layer}{d}{(#1)}
\define@cmdkey[bonak]{painting}{D}{(#1)}
\define@cmdkey[bonak]{painting}{E}{(#1)}
\define@cmdkey[bonak]{painting}{d}{(#1)}
\define@cmdkey[bonak]{fullframe}{D}{(#1)}

\define@cmdkey[bonak]{restrframe}{D}{(#1)}
\define@cmdkey[bonak]{restrframe}{d}{(#1)}
\define@cmdkey[bonak]{restrlayer}{D}{(#1)}
\define@cmdkey[bonak]{restrlayer}{d}{(#1)}
\define@cmdkey[bonak]{restrlayer}{l}{(#1)}
\define@cmdkey[bonak]{restrpainting}{D}{(#1)}
\define@cmdkey[bonak]{restrpainting}{E}{(#1)}
\define@cmdkey[bonak]{restrpainting}{d}{(#1)}
\define@cmdkey[bonak]{restrpainting}{c}{(#1)}

\define@cmdkey[bonak]{cohframe}{D}{(#1)}
\define@cmdkey[bonak]{cohframe}{d}{(#1)}
\define@cmdkey[bonak]{cohlayer}{D}{(#1)}
\define@cmdkey[bonak]{cohlayer}{d}{(#1)}
\define@cmdkey[bonak]{cohlayer}{l}{(#1)}
\define@cmdkey[bonak]{cohpainting}{D}{(#1)}
\define@cmdkey[bonak]{cohpainting}{E}{(#1)}
\define@cmdkey[bonak]{cohpainting}{d}{(#1)}
\define@cmdkey[bonak]{cohpainting}{c}{(#1)}
\makeatother

\newcommandx{\X}[3][1,2,3]{
  \ensuremath{{\color{carolina}{\nu\textsf{Set}}}_{#1}^{#2}}
  \setkeys*[bonak]{X}{#3}
}

\newcommandx{\prim}[6][2,3,4,5,6]{
  \ensuremath{\mathsf{\color{indian-yellow}{#1}}_{#2}^{
    \joinargs[#3][#4][#5]}}
  \setkeys*[bonak]{#1}{#6}
}

\newcommandx{\restr}[8][2,3,4,5,6,7,8]{
  \ensuremath{\mathsf{\color{russian-green}{restr}}_{
    \joinargs[\mathsf{\color{indian-yellow}{#1}}][#2][#3][#4]}^{\joinargs[#5][#6][#7]}}
  \setkeys*[bonak]{restr#1}{#8}
}

\newcommandx{\coh}[9][2,3,4,5,6,7,8,9]{
  \ensuremath{\mathsf{\color{chestnut}{coh}}_{
    \joinargs[\mathsf{\color{indian-yellow}{#1}}][#2][#3][#4][#5][#6]}^{\joinargs[#7][#8]}}
  \setkeys*[bonak]{coh#1}{#9}
}

\newcommandx{\cohtwo}[9][2,3,4,5,6,7,8,9]{
  \ensuremath{\mathsf{\color{chestnut}{coh2}}_{
    \joinargs[\mathsf{\color{indian-yellow}{#1}}][#2][#3][#4][#5][#6][#7]}^{\joinargs[#8]}}
  \setkeys*[bonak]{coh#1}{#9}
}

\newcommand{\DeltaPlus}{\ensuremath{\boldsymbol{\Delta}_+}}
\newcommand{\SSSet}{\ensuremath{\Set_{\DeltaPlus}}}
\newcommand{\Cube}{\ensuremath{\boldsymbol{{\square}}}}
\newcommand{\CSet}{\ensuremath{\Set_{\Cube}}}

\renewcommandx{\U}[1][1=]{\ensuremath{\mathsf{\color{spanish-blue}{HSet}}_{#1}}}
\newcommandx{\HGpd}[1][1=]{\ensuremath{\mathsf{\color{spanish-blue}{HGpd}}_{#1}}}
\newcommand{\Type}{\ensuremath{\mathsf{\color{spanish-blue}{Type}}}}
\newcommand{\SProp}{\ensuremath{\mathsf{\color{spanish-blue}{SProp}}}}

\newcommand{\unittype}{\ensuremath{\mathsf{unit}}}

\newcommand{\unitpoint}{\ensuremath{\ast}}

\newcommand{\defeq}{\ensuremath{\triangleq}}
\newcommand{\refl}{\ensuremath{\mathsf{refl}}}

\newcommand{\tl}{\ensuremath{\mathsf{tl}}}
\newcommand{\hd}{\ensuremath{\mathsf{hd}}}
\newcommand{\imp}{\rightarrow}
\newcommand{\overright}[1]{\overrightarrow{#1}}
\renewcommand{\D}{D}
\newcommand{\hdD}{D.1}
\newcommand{\tlD}{D.2}
\renewcommand{\d}{d}
\renewcommand{\E}{E}
\newcommand{\ap}{\mathsf{ap}\;}
\renewcommand{\l}{l}
\renewcommand{\c}{c}
\newcommand{\pair}[2]{#1, #2}
\newcommand{\Dom}{\textsf{Dom}}
\newcommand{\UIP}{\textsf{UIP}}

\newcommandx{\framep}[2][1,2]{\prim{frame}[][#1][#2][][]}
\newcommandx{\layer}[2][1,2]{\prim{layer}[][#1][#2][][]}
\newcommandx{\painting}[2][1,2]{\prim{painting}[][#1][#2][][]}
\newcommandx{\restrf}[4][1,2,3,4]{\restr{frame}[][#3][#4][#1][#2][][]}
\newcommandx{\restrl}[4][1,2,3,4]{\restr{layer}[][#3][#4][#1][#2][][]}
\newcommandx{\restrc}[4][1,2,3,4]{\restr{painting}[][#3][#4][#1][#2][][]}
\newcommandx{\cohf}{\coh{frame}[][][][][][][][]}
\newcommandx{\cohl}{\coh{layer}[][][][][][][][]}
\newcommandx{\cohc}{\coh{painting}[][][][][][][][]}
\newcommandx{\coht}{\cohtwo{frame}[][][][][][][][]}
\newcommandx{\fullframe}[1][1]{\prim{fullframe}[][#1][][][]}

\newcommand \seqr[3]
  {\shortstack{$#2$ \\ \mbox{}\\
                   \mbox{}\hrulefill\mbox{}\\ \mbox{}\\ $#3$} \raisebox{2ex}{$\;\;\mbox{$#1$}$}}

\newcommand{\kstar}{{\star}}

\DeclareCaptionFormat{plain}{#1: #3}
\newcolumntype{Y}{>{\centering\arraybackslash}X}
\NewEnviron{eqntable}[1]{
  \begin{table}[h]
  \small
    \begin{tabularx}{0.94\linewidth}{
      @{}
      >{$}l<{$}
      >{$}c<{$}
      >{$}c<{$}
      >{$}Y<{$}
      @{}}
      \toprule
      \BODY
      \bottomrule
    \end{tabularx}
    \caption{#1}
  \end{table}
}

\def\graymidrule{\arrayrulecolor{gray30}\midrule\arrayrulecolor{gray65}}

\NewDocumentCommand{\eqnline}{m m m m}{#1 & #2 & #3 & #4 \\}
\newcommandx*{\mc}[1]{\multicolumn{4}{c}{\emph{#1}} \\\\}

\newcommandx*{\eqnarg}[3]{\ifinmask[bonak]{#1}[#2]{\{#2:#3\}}{(#2:#3)}}

\color{gray65}

\begin{document}
\title{A parametricity-based formalization of semi-simplicial and semi-cubical sets}
\date{}
\author{
  \textcolor{gray80}{Hugo Herbelin} \\
  \itshape \textcolor{gray80}{Université Paris-Cité, Inria, CNRS, IRIF, Paris} \\
  \ttfamily \textcolor{gray80}{Hugo.Herbelin@inria.fr}
  \and
  \textcolor{gray80}{Ramkumar Ramachandra} \\
  \itshape \textcolor{gray80}{Université Paris-Cité (2020-2022), Unaffiliated} \\
  \ttfamily \textcolor{gray80}{r@artagnon.com}
}
\maketitle
\begin{abstract}
  Semi-simplicial and semi-cubical sets are commonly defined as presheaves over respectively, the semi-simplex or semi-cube category. Homotopy Type Theory then popularized an alternative definition, where the set of $n$-simplices or $n$-cubes are instead regrouped into the families of the fibres over their faces, leading to a characterization we call \emph{indexed}. Moreover, it is known that semi-simplicial and semi-cubical sets are related to iterated Reynolds parametricity, respectively in its unary and binary variants. We exploit this correspondence to develop an original uniform indexed definition of both augmented semi-simplicial and semi-cubical sets, and fully formalize it in Coq.
\end{abstract}

\section{Introduction}
\subsection*{Fibred vs indexed presentation of semi-simplicial and semi-cubical sets}
A family of sets can commonly be represented in two ways: as a family properly speaking, indexed by the elements of a given set $S$, or as a set $T$ together with a map from $T$ to $S$, which specifies for each element of $T$ its dependency on $S$. In the former case, we call it an \emph{indexed} presentation. In the latter case, the set associated to a given element of $S$ is the fibre of this element, so we call it a \emph{fibred} presentation. The two presentations are equivalent and the equivalence can be phrased concisely in the language of homotopy type theory~\cite{hottbook} as the fibred/indexed equivalence\footnote{In an informal discussion, alternative nomenclatures were proposed: fibration/family equivalence and unbundled/bundled equivalence. The fibred/indexed nomenclature echoes the Grothendieck construction of fibred categories from indexed categories. The most elementary instance of the equivalence, with $\Type$ instead of $\U$, is sometimes called ``Grothendieck construction for dummies'', and its proof requires univalence~\cite{hottbook}.}.
\begin{equation*}
  \mbox{(fibred)}\qquad(\Sigma T: \U. (T \rightarrow S)) ~\simeq~ (S \rightarrow \U)\qquad \mbox{(indexed)}
\end{equation*}
Here, $\U$ represents in homotopy type theory the subset of types within a given universe where equality of any two elements has at most one proof.

A \emph{presheaf} on an category is a family of sets indexed by the object of the category with maps indexed by the morphisms. As such, it lives on the indexed side of the equivalence, contrasting with the fibred side, where we have \emph{discrete Grothendieck fibrations}~\cite{LoregianRiehl20}. However, there are situations where a presheaf can also be seen as living on the fibred side of the equivalence. This happens when the indexing category is \emph{direct}, or has a downwards-well-founded collection of non-identity morphisms. Let us consider, for instance, the case of a semi-cubical set~\cite{grandis03,buchholtz17} presented with $2n$ face maps from the set of $n$-cubes to the set of $(n-1)$-cubes. Formulated in type theory, the corresponding presheaf definition of a semi-cubical set prescribes a family of sets and face maps between them as follows.

\begin{equation*}
  \begin{tikzcd}
    X_0: \U & X_1: \U \arrow[l, "\partial^L" description, shift left=2] \arrow[l, "\partial^R" description, shift right=2] & X_2: \U \arrow[l, "\partial^{L\kstar}" description, shift left=6] \arrow[l, "\partial^{R\kstar}" description, shift left=2] \arrow[l, "\partial^{\kstar L}" description, shift right=2] \arrow[l, "\partial^{\kstar R}" description, shift right=6] & \ldots
  \end{tikzcd}
\end{equation*}
up to cubical faces identities. Here, $X_1$ can be seen as a family over $X_0 \times X_0$, and $X_2$ can be seen as a family over $X_1 \times X_1 \times X_1 \times X_1$, in the fibred presentation, together with coherence conditions between the $X_1$ seen as families over $X_0 \times X_0$. This suggests an alternative indexed presentation of the presheaf as a stratified sequence of families indexed by families of lower rank, taking into account those coherence conditions to prevent duplications. Formulated in type theory, it takes the form:
\begin{equation*}
  \begin{array}{lll}
    X_0 & :               \U                                                                          \\
    X_1 & :               X_0 \times X_0 \rightarrow  \U                                              \\
    X_2 & : \Pi a b c d.\,  X_1(a,b) \times X_1 (c,d) \times X_1(a,c) \times X_1 (b,d) \rightarrow \U \\
    \ldots
  \end{array}
\end{equation*}

The idea for such an indexed presentation of presheaves over a direct category was mentioned at the Univalent Foundations year in the context of defining semi-simplicial types\footnote{\href{https://ncatlab.org/nlab/show/semi-simplicial+types+in+homotopy+type+theory}{ncatlab.org/nlab/show/semi-simplicial+types+in+homotopy+type+theory}}. A few constructions have been proposed since then. The first construction by \cite{voevodsky12} relies on the presentation of semi-simplicial sets as a presheaf over increasing injective maps between finite ordinals. The second, by \cite{herbelin15}\footnote{In hindsight, the title of the paper ``A dependently-typed construction of semi-simplicial types'' is somewhat confusing: it implicitly claimed to construct semi-simplicial types, but the construction was done in a type theory with Uniqueness of Identity Proofs. Consequently, what was really obtained was an indexed presentation of semi-simplicial sets. The confusion was however, common at the time.} formalized in the Coq proof assistant, relies on the presentation of semi-simplicial sets as a presheaf over face maps. Another by \cite{part15} formalized in an emulation of logic-enriched homotopy type theory in the Plastic proof assistant, and yet another by \cite{altenkirch16} formalized in an emulation of a two-level type theory in the Agda proof assistant\footnote{\href{https://github.com/nicolaikraus/HoTT-Agda/blob/master/nicolai/SemiSimp/SStypes.agda}{github.com/nicolaikraus/HoTT-Agda/blob/master/nicolai/SemiSimp/SStypes.agda}}, rely, like in Voevodsky, on the presentation of the semi-simplicial category from increasing injective maps between finite ordinals. The latter constructions, besides being stated as providing semi-simplicial types (thanks to an extension of the type theory), are particularly concise, taking advantage of a definition of increasing injective maps between finite ordinals as type-theoretic functions to inherit the associativity of composition directly from it holding in type theory. This contrasts with the combinatorial construction in \cite{herbelin15} where equations over face maps have to be proved by induction.

By taking the sum of each component of an indexed presentation over the indexing set of this component, one obtains back a presheaf in the ordinary sense that has a property of \emph{Reedy fibrancy}, that is whose morphisms are projections in the set-theoretic sense. Such Reedy fibrant presheaves over a direct category have been studied in e.g. \cite{shulman15}, \cite{kraus17} and \cite{annenkovCK17,AnnenkovCKS2023}, presenting generic constructions over such presheaves.

The indexed definition of a presheaf over a direct category is technically more involved than the presheaf definition, as it requires hard-wiring in the structure the dependencies between elements of the sets of the presheaf, including the coherence conditions between these dependencies, such as taking the $i$-th face of the $j$-th face of a $n$-simplex being the same as taking the $(j-1)$-th face of the $i$-th face (when $j>i$). However, exhibiting a concrete instance of a presheaf in indexed form only requires providing the families, since the responsibility of defining maps and showing the coherence conditions is already accounted for in the definition of the structure.

\subsection*{Reynolds parametricity and its unary and binary variants}
In the context of functional programming, Reynolds parametricity~\cite{reynolds83} interprets types as relations characterizing the observational behaviour of programs of this type. More generally, families over a product of sets, or \emph{correspondences}, can be used in place of relations. Parametricity can then be iterated, and relying on the fibred presentation of correspondences as spans, it has been noted that iterated Reynolds parametricity has the structure of a cubical set~\cite{altenkirch15,moulin16,johann17,moeneclaey21,moeneclaey22phd}. We obtain a \emph{unary} variant of Reynolds \emph{binary} parametricity by using predicates or families instead of relations or correspondences, and this is a form of realizability~\cite{bernardy12,lasson12,moulin16}. Cubical set models which differ only by the arity one~\cite{bernardy15} or two~\cite{bezem13} were introduced, and this led to a general notion of affine $\nu$-ary cubes in \cite{nuytsdevriese24}. In parallel, it has been noted that iterated unary parametricity has the structure of an augmented simplicial set\footnote{Private communication with Hugo Moeneclaey and Thorsten Altenkirch.}. This suggests that the definition of augmented semi-simplicial sets and semi-cubical sets can in turn be seen as particular instances of the restriction of Nuyts-Devriese's $\nu$-ary cubes to only faces, which we call $\nu$-sets, of presheaves over a $\nu$-semi-shape category made of words of some cardinal $\nu+1$, where $\nu=1$ gives augmented semi-simplicial sets and $\nu=2$ gives semi-cubical sets.

\subsection*{Contribution}
The main contribution of the paper is to describe the details of a recipe that uniformly characterizes unary and binary iterated parametricity in indexed form, and to derive from it a new indexed presentation, called indexed \emph{$\nu$-sets}, of augmented semi-simplicial and semi-cubical sets.

Our work is a step in the direction of the program initiated in \cite{altenkirch15} to develop parametricity-based models of parametric type theory~\cite{bernardy15,nuyts17,cavallo19} and cubical type theory~\cite{bezem13,cohen16,angiuli21}, which are closer to the syntax of type theory, and are likely to better reflect the definitional properties of type theory than presheaf-based cubical sets would. For example, consider the loss of definitional properties when interpreting ``indexed'' dependent types of type theory as ``fibrations'' in models such as locally cartesian closed categories~\cite{curien14}.

Our construction has the unique property of reflecting the structure of parametricity and of yielding both augmented semi-simplicial and semi-cubical sets from the same construction. The approach taken in \cite{part15} and \cite{altenkirch16} takes benefit of the definitional compositionality of increasing injective maps, but we do not see how they could be generalized to yield semi-cubical sets.

Our mechanization can be found at \href{https://github.com/artagnon/bonak}{github.com/artagnon/bonak}. The construction was conceived in Summer 2019, and the mechanization began in late 2019. A sketch of the construction was presented at the 2020 HoTT-UF workshop, and the completion of the mechanization was reported at the TYPES 2022 conference.

\section{Semi-simplicial and semi-cubical sets\label{sec:nu}}
In this section, we generalize semi-simplicial and semi-cubical sets to $\nu$-sets, subsuming the earlier definitions. We start with some introductory material on semi-simplicial and semi-cubical sets.

\subsection*{Augmented semi-simplicial sets}
Augmented semi-simplicial sets are defined similarly to semi-simplicial sets, except that the connected components are additionally dependent on a ``colour''. Conversely, semi-simplicial sets can be seen as augmented semi-simplicial sets over the singleton set of a fixed colour. Let us associate dimension $-1$ to colours; then, points are dimension $0$, lines are dimension $1$, and so on.

Ordinary semi-simplicial sets are presheaves over the semi-simplex category. Augmented semi-simplicial sets are presheaves over \DeltaPlus. There are different ways to define \DeltaPlus, up to equivalence, and we use a definition that can be extended to semi-cubical sets in a straightforward manner. In particular, we start numbering objects from $0$ instead of $-1$ so that there is a shift by one compared to the standard numbering of augmented semi-simplicial sets.

\begin{notation}[Finite sequences]
  We denote finite sequences by $i_1 \ldots i_n$ for $i_j$ ranging over some domain. The empty sequence is written $\epsilon$.
\end{notation}

\begin{definition}[$\DeltaPlus$]
  The definition of $\DeltaPlus$ is shown below. Note that, if $g \circ f$ is well-defined, then the length of $f$ is less than or equal to that of $g$. It can be shown that composition is associative and that $\id$ is neutral.
  \begin{align*}
    \Obj_{\DeltaPlus}       & := \Nat                                                                         \\
    \Hom_{\DeltaPlus}(p, n) & := \{l \in \{0, \kstar\}^n \mid \text{number of $\kstar$ in $l = p$}\}          \\
    g \circ f               & :=
    \begin{cases}
      f                & \text{if $g = \epsilon$}                                           \\
      0\,(g' \circ f)  & \text{if $g = 0\,g'$}                                              \\
      a\,(g' \circ f') & \text{if $g = \kstar\,g'$, $f = a\,f'$, where $a = 0$ or $\kstar$} \\
    \end{cases}                     \\
    \id                     & := \kstar \ldots \kstar \text{ $n$ times for $\id \in \Hom_{\DeltaPlus}(n, n)$}
  \end{align*}
\end{definition}

\begin{definition}[$\SSSet$]
  We define the category of augmented semi-simplicial sets as the functor category:
  \begin{equation*}
    \SSSet := \Set^{\DeltaPlus^{\mathsf{op}}}
  \end{equation*}
\end{definition}

To provide examples, we define the standard augmented $n$-semi-simplex, taking into account the shift by one in the numbering.

\begin{definition}[$\DeltaPlus^n$]
  The standard augmented $(n-1)$-semi-simplex $\DeltaPlus^{n-1}$ is defined as the Yoneda embedding of $n \in \Obj(\DeltaPlus)$:
  \begin{align*}
    \DeltaPlus^{n-1}    & : \SSSet                   \\
    \DeltaPlus^{n-1}(p) & := \Hom(p, n)              \\
    \DeltaPlus^{n-1}(f) & := \lambda g .\, g \circ f
  \end{align*}
\end{definition}

The standard augmented $(-1)$-semi-simplex is a singleton made of one colour (in this case, black). Standard augmented $n$-semi-simplices for $n \geq 0$ have a geometric interpretation, and we illustrate them for dimensions $0$, $1$, and $2$.

\begin{example}[$\DeltaPlus^0$]
  The standard augmented $0$-semi-simplex can be pictured as a point, coloured black, corresponding to the unique morphism in $\Hom(0, 1)$. This point is the identity in $\Hom(1, 1)$; it is hence shown as a singleton $\kstar$.
  \begin{equation*}
    \begin{tikzcd}
      \kstar
    \end{tikzcd}
  \end{equation*}
\end{example}

\begin{example}[$\DeltaPlus^1$]
  The standard augmented $1$-semi-simplex is drawn as two points, given by $\Hom(1, 2)$, along with a line connecting them, given by $\Hom(2, 2)$. We use black to denote the unique morphisms in $\Hom(0, 1)$ and $\Hom(0, 2)$.
  \begin{equation*}
    \begin{tikzcd}
      \kstar0 \arrow[r, dash, "\kstar\kstar"] & 0\kstar
    \end{tikzcd}
  \end{equation*}
\end{example}

\begin{example}[$\DeltaPlus^2$]
  $\DeltaPlus^2$ is drawn as three points, given by $\Hom(1, 3)$, three lines connecting them, given by $\Hom(2, 3)$, and a triangular filler given by $\Hom(3, 3)$.
  
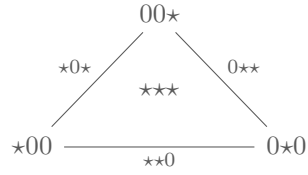
\begin{equation*}
    \begin{tikzcd}
      & |[alias=F]|00\kstar \arrow[ddr, dash, "0\kstar\kstar"] & \\\\
      \kstar00 \arrow[rr, dash, "\kstar\kstar0"{name=T, below}]\arrow[uur, dash, "\kstar0\kstar"] && 0\kstar0 \\
      \arrow[rightarrow, from=F, to=T, phantom, "\kstar\kstar\kstar" description]
    \end{tikzcd}
  \end{equation*}
\end{example}

More generally, the standard augmented $(n + 1)$-semi-simplex can be obtained by taking a copy of the standard augmented $n$-semi-simplex serving as a base, and gluing on top of it another copy lifted by one dimension. In the second copy, the colour becomes an extra point, the points become lines connecting the points of the base to the extra point, and so on. In particular, the components of the base are those of the standard augmented $n$-semi-simplex postfixed by $0$ while the components of the lifted copy are postfixed by $\kstar$. Note that the components may be oriented by letting each $n$-dimensional component point to the $(n-1)$-dimensional component obtained by replacing the leftmost $\kstar$ of the $n$-dimensional component with $0$.

\subsection*{Semi-cubical sets}
Semi-cubical sets are defined like augmented semi-simplicial sets except that $\DeltaPlus$ is replaced by $\Cube$ in which we take sequences of $L$, $R$ and $\kstar$, instead of sequences of $0$ and $\kstar$.

\begin{definition}[\Cube]
  The definition of $\Cube$ is shown below. The symbols $L$ and $R$ indicate opposite faces of a cube.

  \begin{align*}
    \Obj_{\Cube}       & := \Nat                                                                   \\
    \Hom_{\Cube}(p, n) & := \{l \in \{L, R, \kstar\}^n \mid \text{number of $\kstar$ in $l = p$}\} \\
    g \circ f          & :=
    \begin{cases}
      f                & \text{if $g = \epsilon$}                                                 \\
      a\,(g' \circ f)  & \text{if $g = a\,g'$}, \text{where $a = L$ or $R$}                       \\
      a\,(g' \circ f') & \text{if $g = \kstar\,g'$, $f = a\,f'$, where $a = L$, $R$, or $\kstar$} \\
    \end{cases}    \\
    \id                & := \kstar \ldots \kstar \text{ $n$ times}
  \end{align*}

  Again, if $g \circ f$ is well-defined, then the length of $f$ is less than or equal to that of $g$. It can be shown that composition is associative and that $\id$ is neutral.
\end{definition}

\begin{definition}[\CSet]
  We define the category of semi-cubical sets as the functor category:
  \begin{equation*}
    \CSet := \Set^{\op{\Cube}}
  \end{equation*}
\end{definition}

\begin{definition}[$\Cube^n$]
  The standard semi-cube $\Cube^n$ is defined as the Yoneda embedding of $n \in \Obj(\Cube)$:
  \begin{align*}
    \Cube^n    & : \CSet                    \\
    \Cube^n(p) & := \Hom(p, n)              \\
    \Cube^n(f) & := \lambda g .\, g \circ f
  \end{align*}
\end{definition}

Standard $n$-semi-cubes have a geometric interpretation, which we illustrate for dimensions $0$, $1$, and $2$.

\begin{example}[$\Cube^0$]
  $\Cube^0$ is $\Hom(0, 0)$, or the singleton set of the empty sequence:
  \begin{equation*}
    \begin{tikzcd}
      \epsilon
    \end{tikzcd}
  \end{equation*}
\end{example}

\begin{example}[$\Cube^1$]
  $\Cube^1$ consists of two points, given by $\Hom(0, 1)$, and a line, given by $\Hom(1, 1)$.
  \begin{equation*}
    \begin{tikzcd}
      L \arrow[r, "\kstar", dash] & R
    \end{tikzcd}
  \end{equation*}
\end{example}

\begin{example}[$\Cube^2$]
  $\Cube^2$ consists of four points, given by $\Hom(0, 2)$, four lines connecting the four points, given by $\Hom(1, 2)$, and a filler, given by $\Hom(2, 2)$:
  \begin{equation*}
    \begin{tikzcd}
      LR \arrow[r, dash, "\kstar R"{name=F}] \arrow[d, dash, "L\kstar" left] & RR \arrow[d, dash, "R\kstar"] \\
      LL \arrow[r, dash, "\kstar L"{name=T, below}] & RL \\
      \arrow[rightarrow, from=F, to=T, phantom, "\kstar\kstar" description]
    \end{tikzcd}
  \end{equation*}
\end{example}

More generally, the standard $(n + 1)$-semi-cube can be obtained by taking two copies of the standard $n$-semi-cube serving as bottom and top face and connecting them on their border by a prism obtained as a third copy stretched in the new dimension. The bottom and top faces are obtained from the standard $n$-semi-cube by postfixing with respectively $L$ and $R$ while the prism is obtained by postfixing with $\kstar$. Note that the components can be oriented by letting each $n$-dimensional component go from the $(n-1)$-dimensional component obtained by replacing the leftmost $\kstar$ with $L$, to the one obtained by replacing the leftmost $\kstar$ with $R$.

\subsection*{\texorpdfstring{$\nu$}{ν}-sets}
Let us call $\nu$-sets, the generalization of augmented semi-simplicial sets and semi-cubical sets obtained by building on an arbitrary alphabet $\nu$, so that the following holds:

\begin{center}
  \begin{tabularx}{\linewidth}{X|X|X}
    \toprule
    Cardinal of $\nu$ & 1                              & 2                 \\
    \graymidrule
    Interpretation    & Augmented semi-simplicial sets & Semi-cubical sets \\
    \bottomrule
  \end{tabularx}
\end{center}

To obtain this, we extend $\DeltaPlus$ and $\Cube$ in a straightforward manner into a category which we call $\hexagon$.

\begin{definition}[$\hexagon$]
  The definition of $\nu$-semi-shape category is shown below. Note that, if $g \circ f$ is well-defined, then the length of $f$ is less than or equal to that of $g$. It can be shown that composition is associative and that $\id$ is neutral.
  \begin{align*}
    \Obj_{\hexagon}       & := \Nat                                                                          \\
    \Hom_{\hexagon}(p, n) & := \{l \in (\nu \sqcup \{\kstar\})^n \mid \text{number of $\kstar$ in $l = p$}\} \\
    g \circ f             & :=
    \begin{cases}
      f                & \text{if $g = \epsilon$}                                                   \\
      a\,(g' \circ f)  & \text{if $g = a\,g'$}, \text{where $a \in \nu$}                            \\
      a\,(g' \circ f') & \text{if $g = \kstar\,g'$, $f = a\,f'$, where $a \in \nu$ or $a = \kstar$} \\
    \end{cases}            \\
    \id                   & := \kstar \ldots \kstar \text{ $n$ times for $\id \in \Hom_{\hexagon}(n, n)$}
  \end{align*}
\end{definition}

A $\nu$-set is thus a contravariant functor $\phi$ from the $\nu$-semi-shape category to $\Set$ and we call $n$-$\nu$-semi-shape an element of $\phi(n)$. As in the augmented semi-simplicial and semi-cubical cases, the standard $(n + 1)$-$\nu$-semi-shape is obtained by connecting together $\nu$ copies of the standard $n$-$\nu$-semi-shape with an extra copy stretched in the new dimension. We clarify in the next sections, how this process of construction is similar to the parametricity translation developed for functional programming~\cite{reynolds83} and more generally for type theory~\cite{bernardy10,bernardy11,atkey14,bernardy15}.

\section{Type theory\label{sec:tt}}
Martin-Löf's Type theory~(\cite{martinlof75,martinlof84}) is a logical formalism based on the notion of a \emph{type} rather than that of a \emph{set}. It can be seen as a foundation of mathematics alternative to set theory and is the core of several tools for the formalization of mathematics such as Agda, Coq and Lean. In type theory, propositions are types and proofs are programs. Type theory includes \emph{definitional equality}, by which all propositions and proofs are quotiented.

Type theory is a flexible formalism supporting different models. Some models come from homotopy theory, and are based on simplicial sets or related structures~\cite{HofmannStreicher94,kapulkin21,BezemCoquandHuber13,cchm}: in these models, equality is interpreted as paths, and they support the univalence principle stating that equality of types mimics equivalence of types, leading to the development of Homotopy Type Theory~\cite{hottbook}.

Types are organized in a hierarchy of universes written $\Type_m$ for $m$ a natural number. The main types in type theory are the type of dependent pairs, written $\Sigma a : A.\,B(a)$, the type of dependent functions, written $\Pi a : A.\,B(a)$, for $A$ a type and $B(a)$ a type dependent on the inhabitant $a$ of $A$, and the type of propositional equalities, written $t = u$. As a notation, the type of dependent pairs when $B$ is not dependent on $A$ is shortened into $A \times B$ and the type of dependent functions when $B$ is not dependent on $A$ is written $A \rightarrow B$. We assume our type theory to also include a distinguished singleton type, written $\unittype$, and with inhabitant $\unitpoint$, the type of boolean values, and the type of natural numbers. We write $\hd$ and $\tl$ the projections of dependent pairs, and $\refl$ for reflexivity. Logical propositions being types themselves, we use $\Pi$ to represent universal quantification and $\Sigma$ to represent existential quantification. We also assume that our type theory includes a coinductively-defined notion of dependent streams described in the Appendix.

A type-theoretic notion of sets can be recovered in each universe as $\U[m]$, denoting the subtype of $\Type_m$ for which paths are degenerated, using Uniqueness of Identity Proofs (UIP). Technically, this is expressed as a structure equipping a domain $\Dom$ with the property $\UIP$:
\begin{align*}
  \Dom & : \Type_m                                  \\
  \UIP & : \Pi x y: \Dom.\, \Pi p q: x = y.\, p = q
\end{align*}

In $\U[m]$, the following properties hold:
\begin{enumerate}
  \item UIP holds on the unit type, bool type, as well as all types of finite cardinal $\nu$.
  \item UIP propagates to $\Sigma$-types.
  \item UIP propagates to $\Pi$-types, with some additional functional extensionality axioms.
\end{enumerate}

By notation, $\Type$ and $\U$ mean $\Type_m$ and $\U[m]$ at some unspecified universe level $m$.

We are also interested in \emph{extensional} type theory, a type theory with the following reflection rule, where $=$ is propositional equality in some type and $\equiv$ is definitional equality~\cite{martinlof84}:
\begin{equation*}
  \seqr{}{\Gamma \vdash p: t = u}{\Gamma \vdash t \equiv u}
\end{equation*}

Note that the reflection rule implies UIP so that $\U$ and $\Type$ are equivalent in extensional type theory. The rule also implies functional extensionality. Extensional type theory is logically equivalent to intensional type theory extended with UIP and functional extensionality~\cite{HofmannPhd}.

\section{Relating to parametricity\label{sec:rel-param}}
Recall from the introduction, the form taken by the indexed presentation of a semi-cubical set:
\begin{equation*}
  \begin{array}{lll}
    X_0 & :               \U                                                                         \\
    X_1 & :               X_0 \times X_0 \rightarrow \U                                              \\
    X_2 & : \Pi a b c d.\, X_1(a,b) \times X_1 (c,d) \times X_1(a,c) \times X_1 (b,d) \rightarrow \U \\
    \ldots
  \end{array}
\end{equation*}

Here, the process of construction of the type of $X_1$ from that of $X_0$, and of the type of $X_2$ from that of $X_1$, is similar to iteratively applying a binary parametricity translation. The binary parametricity which we consider interprets a closed type $A$ by a family $A_\kstar$ over $A \times A$, and this can be seen as a graph whose vertices are in $A$. Each type constructor is associated with the construction of a graph. To start with, the type of types $\U$ is interpreted as the family of type of families ${\U}_\kstar$, which takes $A_L$ and $A_R$ in $\U$ and returns the type $A_L \times A_R \rightarrow \U$ of families over $A_L$ and $A_R$. Also, for $A$ interpreted by $A_\kstar$ and $B(a)$, for $a:A$, interpreted by $B_\kstar((a_L,a_R),a_\kstar)$ with $a_\kstar:A_\kstar(a_L,a_R)$, a dependent function type $\Pi a: A.\, B(a)$ is interpreted as the graph $(\Pi a: A.\, B(a))_\kstar$ that takes two functions $f_L$ and $f_R$ of type $\Pi a: A.\, B(a)$, and expresses that these functions map related arguments in $A$ to related arguments in $B$:
\begin{equation*}
  \begin{array}{l}
    (\Pi a: A.\, B(a))_\kstar(f_L,f_R) \; \defeq \\
    \multicolumn{1}{r}{\qquad\Pi (a_L,a_R): (A \times A).\, \Pi a_\kstar: (A_\kstar(a_L,a_R)\, B_\kstar((a_L,a_R),a_\kstar)(f_L(a_L),f_R(a_R)))}
  \end{array}
\end{equation*}

Similarly, a product type $A \times B$ is interpreted as the graph $(A \times B)_\kstar$ that relates two tuples $(a_L,b_L)$ and $(a_R,b_R)$ in $A \times B$ as follows:
\begin{align*}
  (A \times B)_\kstar((a_L,b_L),(a_R,b_R)) \; \defeq \; A_\kstar(a_L,a_R) \times B_\kstar(b_L,b_R)
\end{align*}

In particular, for $X: \U$, applying our parametricity translation is about associating to $X$ an inhabitant $X_\kstar$ of ${\U}_\kstar(X,X)$ i.e. of $X \times X \rightarrow \U$. In turn, applying the translation again to $X_\kstar: X \times X \rightarrow \U$ is about associating to $X_\kstar$ an inhabitant $X_{\kstar\kstar}$ of $(X \times X \rightarrow \U)_\kstar(X_\kstar,X_\kstar)$ i.e. of:
\begin{align*}
  \Pi ((x_{LL},x_{LR}),(x_{RL},x_{RR})): ((X \times X) \times (X \times X)). \\
  (X_\kstar(x_{LL},x_{LR}) \times X_\kstar(x_{RL},x_{RR})
  \rightarrow X_\kstar(x_{LL}, x_{RL}) \times X_\kstar(x_{LR},x_{RR})  \rightarrow \U)
\end{align*}
which hints us at how the sequence $X_0$, $X_1$, $X_2$ can be seen as a sequence of inhabitants of the iteration of the composition of binary parametricity with the diagonal on types and type families, applied to an initial $X: \U$:
\begin{equation*}
  \begin{array}{lllll}
    X_0 & \defeq & X                & : & \U                                             \\
    X_1 & \defeq & X_\kstar         & : & {\U}_\kstar(X,X)                               \\
    X_2 & \defeq & X_{\kstar\kstar} & : & ({\U}_\kstar(X,X))_{\kstar}(X_\kstar,X_\kstar) \\
    \ldots
  \end{array}
\end{equation*}

This tells us how the informal type given to $X_2$ in the previous section could be rephrased so that it comes as the instance of a general recipe characterizing the type of all $X_i$.

Notice, however, that the recipe obtained so far, $X_{n + 1}: ({S_n})_\kstar(X_n,X_n)$ for $X_n: S_n$, applies parametricity on the \emph{syntax} of the type of $X_n$. It does not directly yield a characterization of $S_n$ as a function from $n$. Reformulating the recipe as an explicit recursive construction, without requiring an interpretation of the syntax of types, is the main outcome of this work, together with the mechanization and the uniform treatment of augmented semi-simplicial and semi-cubical sets by means of the generalization to $\nu$-sets.

\section{Our construction}
In this section, we describe our parametricity-based construction of $\nu$-sets in indexed form at two levels of formality.

Sections~\ref{sec:ett} and~\ref{sec:itt} describe the construction at an informal level of discourse:
\begin{enumerate}
  \item In~\ref{sec:ett}, we present it in informal extensional type theory where equational reasoning is left implicit, and we give an intuition for the construction in~\ref{sec:intuition}.
  \item While reasoning in extensional type theory is similar to reasoning in set theory regarding how equality is handled, extensional type theory has two limitations. The first limitation is that it enforces the principle of Uniqueness of Identity Proofs and this is inconsistent with the Univalence principle, thus making it inexpressible in Homotopy Type Theory. The second limitation is that we want the construction to be formalizable in the Coq proof assistant whose underlying type theory is intensional. Section~\ref{sec:itt} thus rephrases the construction in (informal) intensional type theory. Since $\nu$-sets are $0$-truncated types, we compensate for the absence of UIP by assuming a ``local UIP'', requiring types to be \U.
\end{enumerate}

Sections~\ref{sec:wf},~\ref{sec:le}, and~\ref{sec:eqproperties} describe additional issues to be addressed in order to get a fully formal construction:
\begin{enumerate}
  \item The well-foundedness of the induction requires a special termination evidence which will be discussed in section~\ref{sec:wf}.
  \item The construction is indexed over integers and holds under some constraints on the range of these integers. There is a standard formalization dilemma in this kind of situation: either the constraints on the range are embedded in the construction so that the construction makes sense only on the corresponding range, or the construction is made first on a more general domain than needed but restricted to a smaller domain at the time of use. We adopted the former approach, requiring the construction to be dependent on proofs of inequalities on natural numbers. We discuss how we deal with such dependencies in section~\ref{sec:le}.
  \item A number of standard groupoid properties of equality as well as type isomorphisms have been left implicit in the informal definition. This is discussed in section~\ref{sec:eqproperties}.
\end{enumerate}

\subsection{The construction in informal type theory\label{sec:ett}}
%
\appendmask[bonak]{layer}[D]
\appendmask[bonak]{painting}[D]
\appendmask[bonak]{restrframe}[D]
\appendmask[bonak]{restrlayer}[D, d]
\appendmask[bonak]{restrpainting}[D, d]
\appendmask[bonak]{cohframe}[D]
\appendmask[bonak]{cohlayer}[D, d]
\appendmask[bonak]{cohpainting}[D, d]

\newcommandx{\Xp}[1]{\X[#1][][]}
\newcommandx{\Xto}[3][3=]{\X[#1][<#2][#3]}
\newcommandx{\Xcomp}[3][3=]{\X[#1][=#2][#3]}
\newcommandx{\Xfrom}[3][3=]{\X[#1][\geq#2][#3]}

\renewcommandx{\framep}[5][1,2,3,4,5]{\prim{frame}[][#2][#3][#4][#5]}
\renewcommandx{\layer}[5][1,2,3,4,5]{\prim{layer}[][#2][#3][#4][#5]}
\renewcommandx{\painting}[5][1,2,3,4,5]{\prim{painting}[][#2][#3][#4][#5]}

\renewcommandx{\restrf}[7][1,2,3,4,5,6,7]{\restr{frame}[][#2][#3][#4][#5][#6][#7]}
\renewcommandx{\restrl}[7][1,2,3,4,5,6,7]{\restr{layer}[][#2][#3][#4][#5][#6][#7]}
\renewcommandx{\restrc}[7][1,2,3,4,5,6,7]{\restr{painting}[][#2][#3][#4][#5][#6][#7]}

\renewcommandx{\cohf}[9][1,2,3,4,5,6,7,8,9]{\coh{frame}[][#2][#3][][][][][#9]}
\renewcommandx{\cohl}[9][1,2,3,4,5,6,7,8,9]{\coh{layer}[][#2][#3][][][][][#9]}
\renewcommandx{\cohc}[9][1,2,3,4,5,6,7,8,9]{\coh{painting}[][#2][#3][][][][][#9]}

A $\nu$-set in indexed form is a sequence of families of $\U$, that is $\U[m]$ for some universe level $m$. We call such sequence a $\nu$-set at level $m$, whose type thus lives in $\U[m+1]$.

Table~\ref{tab:coind} describes the type of a $\nu$-set at level $m$ as a dependent stream of type families representing the limit of $n$-truncated $\nu$-sets. Using the notations of Section~\ref{sec:tt} and of the Appendix, the recursive equation $\Xfrom{m}{n}{D} \,\defeq\, \Sigma R: \Xcomp{m}{n}[D=\D].\, \Xfrom{m}{n+1}[D=\pair{D}{R}]$ from the table formally corresponds to the dependent stream given by $Stream_{\Sigma n.\, \Xto{m}{n},\, \lambda (n, \D).\, \Xcomp{m}{n}[D=\D],\,\lambda ((n, D),R).(n+1,(\pair{D}{R}))}(n, \D)$. Therefore, $\Xfrom{m}{n}$ denotes an infinite sequence $X_{n}, X_{n+1}, \ldots$ dependent on a $(<n)$-truncated $\nu$-set, $\Xto{m}{n}$, so that, when $n$ is $0$, it denotes a full $\nu$-set, written $\Xp{m}$. This is made possible because the $(<0)$-truncated $\nu$-set, $\Xto{m}{0}$, is degenerated: it is an empty family, and there is thus only one $(<0)$-truncated $\nu$-set, namely the canonical inhabitant $\kstar$ of $\unittype$.

The definition of the type of a $n$-truncated $\nu$-set is in turn described in table~\ref{tab:core}. In the infinite sequence of type families representing a $\nu$-set, the $n$-th component is a type dependent over a $\fullframe$. It is recursively defined in table~\ref{tab:frames}, using the auxiliary definitions of $\framep$, $\layer$ and $\painting$. A $\fullframe$ describes a boundary of a standard form (simplex, cube), which we decompose into $\layer$, and a $\painting$ corresponds to a filled frame. Notice that the type $\layer$ relies on an operator of frame restriction $\restrf$ which is defined in table~\ref{tab:faces}, and this restriction operator is in turn defined using auxiliary definitions $\restrl$ and $\restrc$.

\def\lab{tab:coind}
\input{tab-coind.tex}

\renewcommandx{\fullframe}[3][1,2,3]{\prim{fullframe}[#1][#2][][][#3]}

\def\lab{tab:core}
\input{tab-core.tex}

\renewcommandx{\fullframe}[3][1,2,3]{\prim{fullframe}[][#2][][][#3]}
\def\lab{tab:frames}
\input{tab-frames.tex}

\renewcommandx{\Xto}[3][3]{\X[][<#2][#3]}
\renewcommandx{\Xcomp}[3][3]{\X[][=#2][#3]}
\renewcommandx{\Xfrom}[3][3]{\X[][\geq#2][#3]}

\renewcommandx{\cohf}[9][1,2,3,4,5,6,7,8,9]{\coh{frame}[][#2][#3][][][#6,#7][#8][#9]}

\def\lab{tab:faces}
\input{tab-faces-ett.tex}

\renewcommandx{\restrf}[7][1,2,3,4,5,6,7]{\restr{frame}[][#2][#3][#4][#5][#6][#7]}
\renewcommandx{\restrl}[7][1,2,3,4,5,6,7]{\restr{layer}[][#2][#3][#4][#5][#6][#7]}
\renewcommandx{\restrc}[7][1,2,3,4,5,6,7]{\restr{painting}[][#2][#3][#4][#5][#6][#7]}

\renewcommandx{\framep}[5][1,2,3,4,5]{\prim{frame}[][][][][#5]}
\renewcommandx{\layer}[5][1,2,3,4,5]{\prim{layer}[][][][][#5]}
\renewcommandx{\painting}[5][1,2,3,4,5]{\prim{painting}[][][][][#5]}

\renewcommandx{\cohf}[9][1,2,3,4,5,6,7,8,9]{\coh{frame}[][#2][#3][][][][][#9]}

\renewcommandx{\cohf}[9][1,2,3,4,5,6,7,8,9]{\coh{frame}[][#2][#3][#4][#5][#6,#7][#8][#9]}
\renewcommandx{\cohl}[9][1,2,3,4,5,6,7,8,9]{\coh{layer}[][#2][#3][#4][#5][#6,#7][#8][#9]}
\renewcommandx{\cohc}[9][1,2,3,4,5,6,7,8,9]{\coh{painting}[][#2][#3][#4][#5][#6,#7][#8][#9]}
\renewcommandx{\coht}[9][1,2,3,4,5,6,7,8,9]{\cohtwo{frame}[][#2][#3][#4][#5][#6][#7,#8][#9]}

\def\lab{tab:coh}
\input{tab-coh-ett.tex}

Notably, the definition of $\restrl$ relies on an equality expressing the commutation of the composition of two $\restrf$. The proof of this commutation is worth being made explicit, which we do in table~\ref{tab:coh} using proof-term notations. The proof requires an induction on the dimension and on the structure of $\framep$, $\layer$, and $\painting$. This is what $\cohf$ does using auxiliary proofs $\cohl$ and $\cohc$. Even though it looks independent of the definitions from the other tables, $\cohf$ has to be proved mutually with the definitions of $\framep$, $\layer$, $\painting$, and their corresponding restrictions. More precisely, for a fixed $n$, the block of $\framep$, $\restrf$, and $\cohf$ has to be defined in one go by induction on $p$. Also, each of $\painting$, $\restrc$, and $\cohc$ is built by induction from $p$ to $n$. The $\painting$ block at $n$ relies on the $\framep$ block at $n$, but the converse dependency is only on lower $n$, so this is well-founded. Note that $\layer$, $\restrl$ and $\cohl$ are just abbreviations. The exact way this mutual recursion is eventually formalized is explained in section~\ref{sec:wf}.

Note that for a fixed constant $n$, relying on the evaluation rules of type theory, the coherence conditions degenerate to a reflexivity proof, so that the construction builds an effective sequence of types not mentioning coherences anymore.

\subsection{Intuition for our formal construction\label{sec:intuition}}
\renewcommandx{\fullframe}[1][1]{\prim{fullframe}[][#1][][][]}
\renewcommandx{\framep}[2][1,2]{\prim{frame}[][#1][#2][][]}
\renewcommandx{\layer}[2][1,2]{\prim{layer}[][#1][#2][][]}
\renewcommandx{\painting}[2][1,2]{\prim{painting}[][#1][#2][][]}
\renewcommandx{\restrf}[4][1,2,3,4]{\restr{frame}[][#3][#4][#1][#2][][]}
\renewcommandx{\restrl}[4][1,2,3,4]{\restr{layer}[][#3][#4][#1][#2][][]}
\renewcommandx{\restrc}[4][1,2,3,4]{\restr{painting}[][#3][#4][#1][#2][][]}
\renewcommandx{\cohf}[1][1]{\coh{frame}[][][][][][#1][][]}
\renewcommandx{\cohl}[1][1]{\coh{layer}[][][][][][#1][][]}
\renewcommandx{\cohc}[1][1]{\coh{painting}[][][][][][#1][][]}
\renewcommandx{\coht}{\cohtwo{frame}[][][][][][][][]}

There is a $\fullframe$ for each dimension $n$, written $\fullframe[n]$, and every $X_n$ is uniformly assigned a type of the form $\fullframe[n] \rightarrow \U$. Here, $\fullframe[n]$ is a ``telescope'' collecting all arguments of the type of $X_i$ in section~\ref{sec:rel-param} as a nesting of $\Sigma$-types.

To illustrate how to recursively build $\fullframe[n]$, let us begin by setting $\fullframe[0] \defeq \unittype$, so that the type $\U$ given to $X_0$ in section~\ref{sec:rel-param} can be equivalently formulated as $\unittype \rightarrow \U$. Then, more generally, let each $\fullframe[n]$ consist of $n$ layers, written $\layer[n][p]$ with $p < n$, that we stack in order, starting from $\unittype$, and writing $\framep[n][p]$ for the $p$ first layers of a $\fullframe[n]$, so that $\fullframe[n]$ is $\framep[n][n]$. For instance, $X_1$ is made of one layer, so that it can be written as a $\Sigma$-type of an inhabitant of $\unittype$ and $\layer[1][0]$. Then, $X_2$ is similarly made of two layers.

\begin{small}
  \begin{equation*}
    \begin{array}{ll}
      X_0                                : \underbrace{\unittype}_{\framep[0][0]}                                                    \rightarrow \U \\
      X_1                                : \underbrace{\Sigma \unitpoint: \unittype. \underbrace{\left(
      \begin{array}{c}
          \underbrace{X_0(\unitpoint)}_{\painting[0][0]}
          \\ \times \\
          \underbrace{X_0(\unitpoint)}_{\painting[0][0]}
        \end{array}\right)}_{\layer[1][0]}}_{\framep[1][1]} \rightarrow \U                                                                            \\
      X_2                                : \underbrace{\Sigma a: \underbrace{\left(\Sigma \unitpoint: \unittype. \underbrace{\left(
          \begin{array}{c}
            \underbrace{\Sigma b: \left(
            \hspace{-0.4em}\begin{array}{c}
                             X_0(\unitpoint)
                             \\ \times \\
                             X_0(\unitpoint)
                           \end{array}\hspace{-0.4em}
            \right).\, \underbrace{X_1 \underbrace{(\unitpoint, b)}_{\restrf[2][0][L][0]}}_{\painting[1][1]}}_{\painting[1][0]}
            \\ \times \\
            \underbrace{\Sigma b: \left(
            \hspace{-0.4em}\begin{array}{c}
                             X_0(\unitpoint)
                             \\ \times \\
                             X_0(\unitpoint)
                           \end{array}\hspace{-0.4em}
            \right).\, \underbrace{X_1 \underbrace{(\unitpoint, b)}_{\restrf[2][0][R][0]}}_{\painting[1][1]}}_{\painting[1][0]}
          \end{array}
          \right)}_{\layer[2][0]}\right)}_{\framep[2][1]}.\underbrace{\left(
        \begin{array}{c}
          \underbrace{X_1 \underbrace{\left(a.\hd, \left(
              \hspace{-0.4em}\begin{array}{c}
                               a.\tl.L.\hd.L, \\
                               a.\tl.R.\hd.L
                             \end{array}\hspace{-0.4em}
              \right)\right)}_{\restrf[2][1][L][1]}}_{\painting[1][1]}
          \\ \times \\
          \underbrace{X_1 \underbrace{\left(a.\hd, \left(
              \hspace{-0.4em}\begin{array}{c}
                               a.\tl.L.\hd.R, \\
                               a.\tl.R.\hd.R
                             \end{array}\hspace{-0.4em}
              \right)\right)}_{\restrf[2][1][R][1]}}_{\painting[1][1]}
        \end{array}
        \right)}_{\layer[2][1]}}_{\framep[2][2]}
      \rightarrow \U                                                                                                                                \\
      \ldots
    \end{array}
  \end{equation*}
\end{small}

Let us now illustrate the construction of $\fullframe[3]$, necessary to build the type of $X_3$.

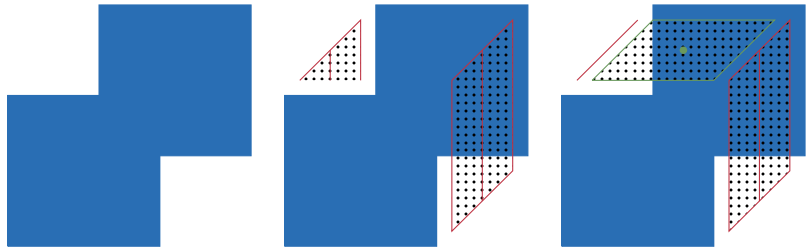
\begin{center}
  \begin{tikzpicture}[scale=2]
    \draw[spanish-blue, fill=spanish-blue] (0, 0) -- (1, 0) -- (1, 1) -- (0, 1) -- (0, 0);
    \draw[spanish-blue, fill=spanish-blue, nearly transparent] (0.6, 1) -- (0.6, 1.6) -- (1.6, 1.6) -- (1.6, 0.6) -- (1, 0.6);
  \end{tikzpicture}
  \;\;
  \begin{tikzpicture}[scale=2]
    \draw[spanish-blue, fill=spanish-blue] (0, 0) -- (1, 0) -- (1, 1) -- (0, 1) -- (0, 0);
    \draw[spanish-blue, fill=spanish-blue, nearly transparent] (0.6, 1) -- (0.6, 1.6) -- (1.6, 1.6) -- (1.6, 0.6) -- (1.0, 0.6);
    \draw[raspeberry, pattern=dots] (1.1, 1.1) -- (1.5, 1.5) -- (1.5, 0.5) -- (1.1, 0.1) -- (1.1, 1.1);
    \draw[raspeberry] (1.3, 1.3) -- (1.3, 0.3);
    \draw[raspeberry, pattern=dots] (0.1, 1.1) -- (0.5, 1.5) -- (0.5, 1.1);
    \draw[raspeberry] (0.3, 1.3) -- (0.3, 1.1);
  \end{tikzpicture}
  \;\;
  \begin{tikzpicture}[scale=2]
    \draw[spanish-blue, fill=spanish-blue] (0, 0) -- (1, 0) -- (1, 1) -- (0, 1) -- (0, 0);
    \draw[spanish-blue, fill=spanish-blue, nearly transparent] (0.6, 1) -- (0.6, 1.6) -- (1.6, 1.6) -- (1.6, 0.6) -- (1.0, 0.6);
    \draw[raspeberry, pattern=dots] (1.1, 1.1) -- (1.5, 1.5) -- (1.5, 0.5) -- (1.1, 0.1) -- (1.1, 1.1);
    \draw[raspeberry] (1.3, 1.3) -- (1.3, 0.3);
    \draw[raspeberry] (0.1, 1.1) -- (0.5, 1.5);
    \draw[russian-green, pattern=dots] (0.2, 1.1) -- (1.0, 1.1) -- (1.4, 1.5) -- (0.6, 1.5) -- (0.2, 1.1);
    \filldraw[russian-green] (0.8, 1.3) circle (0.6pt);
  \end{tikzpicture}
\end{center}

The figure on the left is $\framep[3][1]$, in the middle is $\framep[3][2]$, and on the right is $\framep[3][3]$, which is full. Further, $\framep[3][1]$ is made of one layer, $\layer[3][0]$, shown as the front and back faces (blue boxes), $\framep[3][2]$ is made of one additional layer, $\layer[3][1]$, shown as the left and right faces (red boxes), $\framep[3][3]$ is made of one more layer, $\layer[3][2]$, shown as the top face (green box).

We illustrated here the cubical case, that is $\nu = 2$, but, in general, a $\layer[n][p]$ is a product of $\nu$ $\painting[n - 1][p]$. A $\painting[n][0]$ is a $n$-dimensional object corresponding to a filled $\fullframe[n]$. More generally, a $\painting[n][p]$ is an $n$-dimensional object which has the form of a $\painting[n-p][0]$, thus of $(n-p)$-dimensional form, but shifted and living in dimensions $p$ to $n$. Such $\painting[n][p]$ fills a space framed by a partial $\framep[n][p]$ so that, together, they form a filled $\fullframe[n]$. For instance, in the picture, each of the two $\painting[2][0]$ of $\layer[3][0]$ is a filled blue square, each of the two $\painting[2][1]$ of $\layer[3][1]$ is a line, shown as lines across the left and right faces (red lines), stretched into a partial square filling the partial frames made of respectively, the left and right border of the front-back faces (blue), and each of the two $\painting[2][2]$ of $\layer[3][2]$ is the point shown on the top face (green point), stretched into a partial square filling the full frames made respectively of the upper and lower borders of the front-back and left-right faces (blue and red squares). A $\painting[n][p]$ complements a $\framep[n][p]$ by adding layers needed to form a $\fullframe[n]$ and by filling the resulting $\fullframe[n]$ with an inhabitant of $X_n$. Layers are added from dimension $n$ to dimension $p$, opposite to the order from $0$ to $p$ the $\framep[n][p]$ are built, as shown below.
\begin{equation*}
  \begin{array}{lll}
    \framep[n][p]   & \defeq & \Sigma a_n: (\ldots (\Sigma \unitpoint: \unittype .\, \layer[n][0]) \ldots) .\, \layer[n][p - 1] \\
    \painting[n][p] & \defeq & \Sigma l_p: \layer[n][p]. (\ldots (\Sigma l_n: \layer[n][n - 1] .\, X_n) \ldots)
  \end{array}
\end{equation*}

So far, we have not paid attention to the fact that we have a dependent type, shown as $\Sigma$. To be more precise, note that $\fullframe[n]$ depends on all $X_i$ up to $n - 1$. So, we need to package up $X_i$, for $i < n$, into a nesting of $\Sigma$-types, constituting the type of a $n$-truncated $\nu$-set, which we wrote $\X[][<n][]$. This allows us to give the type $\X[][<n][] \rightarrow \U$ to $\fullframe[n]$. Then, for $D: \X[][<n][]$, representing an initial prefix of $X_0, X_1, \ldots X_{n - 1}$, the indexed set $X_n$ has type $\fullframe[n](D) \rightarrow \U$. Thus, $\framep[n][p]$, $\layer[n][p]$ and $\painting[n][p]$ also depend on $D$. We can then reformulate the previous equation with its dependency on $D$. In particular, $X_n$ is just the last component of $D$, that is $D.\tl$.
\begin{equation*}
  \begin{array}{lll}
    \framep[n][p](D)   & \defeq & \Sigma a_n: (\ldots (\Sigma \unitpoint: \unittype.\, \layer[n][0](D)) \ldots) .\,\layer[n][p - 1](D) \\
    \painting[n][p](D) & \defeq & \Sigma l_p: \layer[n][p](D).\, (\ldots (\Sigma l_n: \layer[n][n - 1](D) .\, D.\tl) \ldots)
  \end{array}
\end{equation*}

An extra refinement arises from the fact that each new layer of a frame has to be glued onto the border of the partial frame built so far. So, each $\layer[n][p]$ has to depend on $\framep[n][p]$. We also need a way to characterize the $\nu$ borders of each $\painting[n-1][p]$ that composes a $\layer[n][p]$, and this is where the restriction $\restrf[n][p][\epsilon][p]$ arrives, for each $\epsilon < \nu$. For instance, in the picture, the left and right faces (red), $\painting[2][1]$, are laid on respectively the left and right borders of the front and back faces (blue), and hence need to depend on $\framep[3][1]$. The left and right borders of the front and back faces are then extracted as $\restrf[2][1][L](D)(d)$ and $\restrf[2][1][R](D)(d)$. We can then refine again the previous equation by showing the dependencies on $d$, as shown below.
\begin{equation*}
  \begin{array}{llr}
    \framep[n][p](D)      & \defeq & \Sigma d: (\ldots (\Sigma \unitpoint: \unittype.\, \layer[n][0](D)(\unitpoint)) \ldots).\, \layer[n][p](D)(d) \\
    \painting[n][p](D)(d) & \defeq & \Sigma l_p: \layer[n][p](D)(d).\, (\ldots (\Sigma l_n: \layer[n][n - 1](D)(d, l_p, \ldots, l_{n - 1}).        \\
                          &        & D.\tl(d, l_p, \ldots, l_n)) \ldots)                                                                           \\
                          &        & \text{where } (d, l_p, \ldots, l_q) \text{ abbreviates } ((\ldots(d, l_p), \ldots), l_q)
  \end{array}
\end{equation*}

When $\nu = 2$, using $L$ and $R$ to represent the sides, the formation of layers from paintings amounts to:
\begin{equation*}
  \begin{array}{llc}
    \layer[n][p](D)(d) & \defeq & \painting[n-1][p](D.\hd)(\restrf[n][p][L][p](d)) \times \painting[n-1][p](D.\hd)(\restrf[n][p][R][p](d))
  \end{array}
\end{equation*}

The operation $\restrf[n][p][\epsilon][q]$ restricts the $p$ first layers of a frame, and the construction is by recursion on the structure of a frame $d$. This necessitates the definitions $\restrl[n][p][\epsilon][q](d)(l)$ and $\restrc[n][p][\epsilon][q](d)(c)$, for $l$ a $\layer$, and $c$ a $\painting$. The key case is $\restrc[n][p][\epsilon][p](d)(c)$, where $c$, a $\painting[n][p]$, necessarily has the form of $((c_L, c_R), \_)$. Here, $\restrc[n][p][L][p]$ picks out $c_L$, a $\painting[n-1][p]$, $\restrc[n][p][R][p]$ picks out the $c_R$, also a $\painting[n-1][p]$, and the last component, shown as $\_$, a $\painting[n][p+1]$, is discarded. There is one last difficulty, which we illustrate by writing down expected and actual types.

Given $c_\omega$ of type
\begin{align*}
  c_\omega & : \painting[n-1][p](D.\hd)(\restrf[n][p][\omega][p](d))
\end{align*}
$\restrl[n][p][\epsilon][q](d)(c_L, c_R)$ produces a layer in which the $\omega$-component has the type
\begin{equation*}
  \painting[n-2][p](D.\hd.\hd)(\restrf[n-1][p][\epsilon][q](\restrf[n][p][\omega][p](d)))
\end{equation*}
while we expect a component of type
\begin{equation*}
  \painting[n-2][p](D.\hd.\hd)(\restrf[n-1][p][\omega][p](\restrf[n][p][\epsilon][q+1](d)))
\end{equation*}

Hence, we need a coherence condition to commute the restrictions. Coherence conditions similar to this necessitate what are shown as, $\cohf$, $\cohl$ and $\cohc$ in table~\ref{tab:coh}. These are by induction on the structure of $\framep$, $\layer$ and $\painting$. Note that, for the construction in intensional type theory, we further need a $2$-dimensional coherence condition, $\coht$, for $\cohl$, which is explained in the next section.

\subsection{From extensional to intensional type theory\label{sec:itt}}
\renewcommandx{\cohf}{\coh{frame}[][][][][][][][]}
In this section, we intend to get rid of the reflection rule and make explicit the equational reasoning step needed to rephrase the construction in intensional type theory. For readability purposes, we make only explicit in this section the key coherence conditions of the construction. Other cases of equality reasoning would have to be made explicit to fully obtain a construction in intensional type theory, but these steps are standard enough to be omitted at this stage. See section~\ref{sec:eqproperties} for the details.

The need for transport along a proof of commutation of $\restrf$ in the definition of $\restrl$ is made explicit in table~\ref{fulltab:faces}, where the arrow over $\cohf$ indicates the direction of rewrite.

\renewcommandx{\framep}[5][1,2,3,4,5]{\prim{frame}[][#2][#3][#4][#5]}
\renewcommandx{\layer}[5][1,2,3,4,5]{\prim{layer}[][#2][#3][#4][#5]}
\renewcommandx{\painting}[5][1,2,3,4,5]{\prim{painting}[][#2][#3][#4][#5]}
\renewcommandx{\restrf}[7][1,2,3,4,5,6,7]{\restr{frame}[][#2][#3][#4][#5][#6][#7]}
\renewcommandx{\restrl}[7][1,2,3,4,5,6,7]{\restr{layer}[][#2][#3][#4][#5][#6][#7]}
\renewcommandx{\restrc}[7][1,2,3,4,5,6,7]{\restr{painting}[][#2][#3][#4][#5][#6][#7]}
\renewcommandx{\cohf}[9][1,2,3,4,5,6,7,8,9]{\coh{frame}[][#2][#3][#4][#5][#6,#7][#8][#9]}
\renewcommandx{\cohl}[9][1,2,3,4,5,6,7,8,9]{\coh{layer}[][#2][#3][#4][#5][#6,#7][#8][#9]}
\renewcommandx{\cohc}[9][1,2,3,4,5,6,7,8,9]{\coh{painting}[][#2][#3][#4][#5][#6,#7][#8][#9]}
\renewcommandx{\coht}[9][1,2,3,4,5,6,7,8,9]{\cohtwo{frame}[][#2][#3][#4][#5][#6][#7,#8][#9]}
\def\lab{fulltab:faces}
\renewcommand{\thetable}{4'}
\input{tab-faces.tex}
\def\lab{fulltab:coh}
\renewcommand{\thetable}{5'}
\input{tab-coh.tex}

\renewcommandx{\cohf}{\coh{frame}[][][][][][][][]}
\renewcommandx{\cohl}{\coh{layer}[][][][][][][][]}
\renewcommandx{\coht}{\cohtwo{frame}[][][][][][][][]}
The proof of $\cohf$ itself requires making explicit several rewrites which were invisible in extensional type theory. The commutation of $\restrl$ lives in a type referring to $\cohf$, so we need a transport along the commutation of $\restrf$ in the statement of $\cohl$. The proof of $\cohl$ is the most involved proof of the construction, as it requires a higher-dimensional coherence condition, $\coht$, whose exact formulation is as follows.
\renewcommandx{\cohf}[9][1,2,3,4,5,6,7,8,9]{\coh{frame}[][#2][#3][#4][#5][#6,#7][#8][#9]}
\renewcommandx{\cohl}[9][1,2,3,4,5,6,7,8,9]{\coh{layer}[][#2][#3][#4][#5][#6,#7][#8][#9]}
\renewcommandx{\cohc}[9][1,2,3,4,5,6,7,8,9]{\coh{painting}[][#2][#3][#4][#5][#6,#7][#8][#9]}
\renewcommandx{\coht}[9][1,2,3,4,5,6,7,8,9]{\cohtwo{frame}[][#2][#3][#4][#5][#6][#7,#8][#9]}
\begin{align*}
  \cohf[m][\omega][\theta][r][p][n-1][p][][d = {\restrf[m][\epsilon][q+2][n][p][][d = \d]}]\;\bullet     \\
  \ap \restrf[m][\omega][r][n-2][p][][]\;(\cohf[m][\epsilon][\theta][q+1][p][n][p][][d = \d])\;\bullet   \\
  \cohf[m][\epsilon][\omega][q][r][n-1][p][][d = {\restrf[m][\theta][p][n][p][][d = \d]}] =              \\
  \ap \restrf[m][\theta][p][n-2][p][][]\;(\cohf[m][\epsilon][\omega][q+1][r+1][n][p][][d = \d])\;\bullet \\
  \cohf[m][\epsilon][\theta][q][p][n-1][p][][d = {\restrf[m][\omega][r+1][n][p][][d = \d]}]\;\bullet     \\
  \ap \restrf[m][\epsilon][q][n-2][p][]\;(\cohf[m][\omega][\theta][r][p][n][p][][d = \d])
\end{align*}
where $\ap$ applies a function on two sides of an equality, and $\bullet$ is transitivity of equality. This property of equality proofs holds in \U, and since our construction is done in \U, the term is trivially discharged.

\renewcommandx{\coht}{\cohtwo{frame}[][][][][][][][]}
\renewcommandx{\restrf}{\restr{frame}[][][][][][][]}
\renewcommandx{\cohf}{\coh{frame}[][][][][][][][]}
\renewcommandx{\cohl}{\coh{layer}[][][][][][][][]}
\renewcommandx{\cohc}{\coh{painting}[][][][][][][][]}

Notice that each $\restrl$ in the type of $\cohl$ is hiding a $\cohf$ rewrite: this makes a sum total of three $\cohf$ rewrites on the left-hand side, and two $\cohf$ rewrites on the right-hand side. In the proof term of $\cohl$, $\cohc$ has one $\cohf$ rewrite on its left-hand side. This, combined with the two terms of the form $\ap \cohf$, matches our expectation of three $\cohf$ on the left-hand side, and two $\cohf$ on the right-hand side. Then, $\coht$ can be seen as expressing the commutation of these $\cohf$ terms.

Finally, let us explain $\cohc$. The base case $p = r$ is the key case of the commutation of $\restrf$, when one of the $\restrc$ collapses, and the remaining equation holds trivially. The case of $p < r$ follows the structure of $\restrc$ by induction.

If we were not working in \U, but in $\HGpd$ we would need to prove one more higher-dimensional coherence, and if we were working in \Type, we would need to prove arbitrarily many higher-dimensional coherences. Here, $\HGpd$ is the subset of types $A$ such that for all $x$ and $y$ in $A$, $x = y$ is in \U. See \cite{herbelin15,altenkirch16,kraus21} for a discussion on the need for recursive higher-dimensional coherence conditions in formulating higher-dimensional structures in type theory.

\subsection{Well-foundedness of the construction\label{sec:wf}}
\renewcommandx{\framep}[2][1,2]{\prim{frame}[][#1][#2][][]}
\renewcommandx{\layer}[2][1,2]{\prim{layer}[][#1][#2][][]}
\renewcommandx{\painting}[2][1,2]{\prim{painting}[][#1][#2][][]}
\renewcommandx{\restrf}[4][1,2,3,4]{\restr{frame}[][#3][#4][#1][#2][][]}
\renewcommandx{\restrl}[4][1,2,3,4]{\restr{layer}[][#3][#4][#1][#2][][]}
\renewcommandx{\restrc}[4][1,2,3,4]{\restr{painting}[][#3][#4][#1][#2][][]}
\renewcommandx{\cohf}[1][1]{\coh{frame}[][][][][][#1][][]}
\renewcommandx{\cohl}[1][1]{\coh{layer}[][][][][][#1][][]}
\renewcommandx{\cohc}[1][1]{\coh{painting}[][][][][][#1][][]}

Since the construction shown in the previous sections is by induction on $n$, and dependencies are on lower $n$ and $p < n$, one would imagine formalizing this using well-founded induction in dependent type theory. We initially tried this approach, and had terms dependent on the proofs of the case distinction that $n' \leq n$ implies $n' < n$ or $n' = n$, but these proofs did not come with enough definitional properties to be usable in practice. Hence, we chose a different route: in practice, since $\restrf[n]$ depends on $\framep[n]$ and $\framep[n-1]$, while $\cohf[n]$ depends on $\framep[n]$, $\framep[n-1]$, and $\framep[n-2]$, we only need to keep track of three consecutive dimensions. Hence, what we build by induction at level $n$, is a structure made not only of the definitions shown in the tables~\ref{tab:frames},~\ref{fulltab:faces}, and~\ref{fulltab:coh}, but also of $\framep$, $\layer$, $\painting$ at levels $n - 1$ and $n - 2$, as well as $\restrf$, $\restrl$, and $\restrc$ at level $n - 1$, together with helper equations.

\subsection{Dependencies in inequality proofs\label{sec:le}}
The entire construction relies on inequalities over natural numbers, and we use two different definitions of $\leq$ addressing different concerns in our formalization. In order to build our first variant, we use an intermediate ``recursive definition'' phrased as:

\begin{minted}{coq}
Fixpoint leR (n m : nat) : SProp :=
match n, m with
| O, _ => STrue
| S n, O => SFalse
| S n, S m => leR n m
end.
\end{minted}

Here, $\SProp$ is a definitionally proof-irrelevant impredicative universe morally\footnote{In Coq, it is however a stand-alone universe unrelated to the universe hierarchy.} living at the bottom of the universe hierarchy~\cite{gilbert19}. By placing the definition in \SProp, we have definitional equality of inequality proofs. However, for the purpose of unification, this definition does not go far enough. Consider the unification problems:

\begin{minted}{coq}
leR_trans ?p leR_refl = ?p
leR_trans leR_refl ?p = ?p
\end{minted}
where \texttt{leR\_trans} is transitivity, \texttt{leR\_refl} is reflexivity, and \texttt{?p} is an existential variable. These two problems definitionally hold in \SProp, but equating them does not solve the existential variable. For unification to be useful in solving existential variables, we present our first variant of $\leq$, which we dub as the ``Yoneda variant'':

\begin{minted}{coq}
Definition leY n m :=
  forall p, leR p n -> leR p m.
\end{minted}

This definition is an improvement over \texttt{leR} since reflexivity is now definitionally the neutral element of transitivity, and associativity of transitivity also holds definitionally. Although it significantly eases our proof, there are some instances where unification is unable to solve the existential variables, and we have to provide them explicitly.

The second variant of $\leq$, the ``inductive variant'', is phrased as:

\begin{minted}{coq}
Inductive leI : nat -> nat -> Type :=
| leI_refl n : n <~ n
| leI_down {n p} : p.+1 <~ n -> p <~ n
where "n <~ m" := (leI n m) : nat_scope.
\end{minted}

Compared to \texttt{leY}, \texttt{leI} has no proof-irrelevance properties. This definition is specially crafted for $\painting$, where we have to reason inductively from $p \leq n$ to $n$. In our usage, we have lemmas \texttt{leY\_of\_leI} and \texttt{leI\_of\_leY} in order to equip \texttt{leY} with the induction scheme of \texttt{leI}. The resulting induction scheme has computational rules holding propositionally.

\subsection{Groupoid properties of equality and basic type isomorphisms\label{sec:eqproperties}}
The construction relies on groupoid properties of equality which are left implicit in table~\ref{fulltab:coh}. The use of the equivalence between $u = v$ and $\Sigma (p:u.\hd = v.\hd). (u.\tl = v.\tl)$ for $u$ and $v$ in a $\Sigma$-type is left implicit in the same table. Also implicit is the use of the equivalence between $f = g$ and $\Pi a: A.\, f(a) = g(a)$ for $f$ and $g$ in $\Pi a: A.\, B$, where it should be recalled that the right-to-left map, or functional extensionality, holds by default in extensional type theory. These have to be made explicit in the formalization.

As a final remark, note that as a consequence of $\eta$-conversion for finite enumerated types, the requirement of functional extensionality disappears when $\nu$ is finite. However, this is a conversion which Coq does not implement, and the alternative would be to replace $\Pi a: \nu.\, B$ by a ``flat'' iterated product $B(1) \times B(2) \times \ldots \times B(\nu)$.

\section{Future work}
The construction could be extended with degeneracies as well as with permutations~\cite{grandis03}. Dependent $\nu$-sets could also be defined, opening the way to construct $\Pi$-types and $\Sigma$-types of $\nu$-sets. A $\nu$-set of $\nu$-sets representing a universe could also be defined as sketched in a talk at the HoTT-UF workshop for the bridge case~(\cite{herbelin-hott-uf}). More generally, we believe these lines of work would eventually provide alternative models to parametric type theories~\cite{nuyts17,cavallo19} where equality of types, now a family rather than the total space of a fibration, is not only definitionally isomorphic to bridges~\cite{bernardy15}, but definitionally the same as bridges.

By equipping the universe construction with a structure of equivalences, as suggested along the lines of \cite{altenkirch15}, we also suspect the construction to be able to serve as a basis for syntactic models of various versions of cubical type theory~\cite{bezem13,cohen16,angiuli21}, saving the detour via the fibred approach inherent to usual presheaf models. In particular, we conjecture being able to justify univalence holding definitionally. Our approach would also firmly ground cubical type theory in iterated parametricity.

Although prior approaches to constructing the indexed presentation of a presheaf over a direct category rely on it being evident by inspection that the fibred and indexed presentations are equivalent, no formal proof has been given, and this is a direction for future work. In our construction, we can check by computation of the first levels that it indeed computes the expected sets.

\bibliographystyle{alpha}
\bibliography{paper}

\newpage
\appendix
\fancyhead[L]{\footnotesize\textcolor{gray80}{{\MakeUppercase{Appendix}}}}
\section*{Appendix: Definition of dependent stream}

\begin{align*}
  \mbox{\emph{type formation}}                                                                                                                                                                                                                                                                                                                                                                                                                                                                                                                                                                                                    \\
  \seqr{}{\Gamma \vdash A : \Type_m \qquad \Gamma, a:A \vdash B(a) : \Type_m \qquad \Gamma, a:A, b:B(a) \vdash f(a,b):A \qquad \Gamma \vdash u:A}
  {\Gamma \vdash Stream_{A,B,f}\, u : \Type_m}                                                                                                                                                                                                                                                                                                                                                                                                                                                                                                                                                                                    \\
  \mbox{\emph{introduction and eliminations}}                                                                                                                                                                                                                                                                                                                                                                                                                                                                                                                                                                                     \\
  \seqr{}{\!\!\!\begin{array}{l}\Gamma \vdash A : \Type_m \qquad \Gamma, a:A \vdash B(a) : \Type_m \qquad \Gamma, a:A, b:B(a) \vdash f(a,b):A \qquad \Gamma \vdash u : A \\ \Gamma, a:A \vdash D(a) : \Type_m \qquad \Gamma \vdash w : D(u) \\ \Gamma, a:A, d:D(a) \vdash v(a,d) : B(a) \qquad \Gamma, a:A, d:D(a) \vdash s(a,d) : D(f(a,v(a,d))) \end{array}}{\Gamma \vdash \textsf{cofix}_{a,d,g}^{u,w}\{this:=v(a,d); next:=g(f(a,v(a,d)),s(a,d))\} : Stream_{A,B,f}\, u}                                                                                                                                                      \\
  \seqr{}{\Gamma \vdash t : Stream_{A,B,f}\, u}{\Gamma \vdash t.this : B(u)}                                                                                                                                                                                                                                                                                                                                                                                                                                                                                                                                                      \\
  \seqr{}{\Gamma \vdash t : Stream_{A,B,f}\, u}{\Gamma \vdash t.next : Stream_{A,B,f}\, f(u,t.this)}                                                                                                                                                                                                                                                                                                                                                                                                                                                                                                                              \\
  \mbox{\emph{computation}}                                                                                                                                                                                                                                                                                                                                                                                                                                                                                                                                                                                                       \\
  \seqr{}{\!\!\!\begin{array}{l}\Gamma \vdash A : \Type_m \qquad \Gamma, a:A \vdash B(a) : \Type_m \qquad \Gamma, a:A, b:B(a) \vdash f(a,b):A \qquad \Gamma \vdash u : A \\ \Gamma, a:A \vdash D(a) : \Type_m \qquad \Gamma \vdash w : D(u) \\ \Gamma, a:A, d:D(a) \vdash v(a,d) : B(a) \qquad \Gamma, a:A, d:D(a) \vdash s(a,d) : D(f(a,v(a,d))) \end{array}}{\Gamma \vdash \textsf{cofix}_{a,d,g}^{u,w}\{this:=v(a,d); next:=g(f(a,v(a,d)),s(a,d))\}.this \equiv v(u,w) : B(u)}                                                                                                                                                 \\
  \seqr{}{\!\!\!\begin{array}{l}\Gamma \vdash A : \Type_m \qquad \Gamma, a:A \vdash B(a) : \Type_m \qquad \Gamma, a:A, b:B(a) \vdash f(a,b):A \qquad \Gamma \vdash u : A \\ \Gamma, a:A \vdash D(a) : \Type_m \qquad \Gamma \vdash w : D(u) \\ \Gamma, a:A, d:D(a) \vdash v(a,d) : B(a) \qquad \Gamma, a:A, d:D(a) \vdash s(a,d) : D(f(a,v(a,d)))\end{array}}{\!\!\!\begin{array}{r}\Gamma \vdash \textsf{cofix}_{a,d,g}^{u,w}\{this:=v(a,d); next:=g(f(a,v(a,d)),s(a,d))\}.next \equiv \\ \textsf{cofix}_{a,d,g}^{f(u,v(u,w)),s(u,w))}\{this:=v(a,d); next:=g(f(a,v(a,d)),s(a,d))\} \\ : Stream_{A,B,f}\,f(u,v(u,w))\end{array}} \\
\end{align*}
where $\textsf{cofix}_{a,d,g}^{u,w}\{this:=v(a,d); next:=g(f(a,v(a,d)),s(a,d))\}$ is a notation for the instantiation on parameter $u$ and internal value $w$ of the corecursive definition of a stream over an arbitrary $a$ generated by a recipe dependent on an arbitrary internal value $d:D(a)$ with first component given by $v(a,d)$ and second component given by $g(f(a,v(a,d)),s(a,d))$ where $g$, typed as $\Gamma, a:A, d:D(a) \vdash g(a,d): Stream_{A,B,f}\, (f(a,d))$, formally represents the recursive call, and where $s(a,d)$ tells how the internal value evolves.
\end{document}

%% file: tab-coind.tex
\providecommand{\lab}{tab:coind}
\begin{eqntable}{Main definition\label{\lab}}
  \eqnline{\Xp{m}}{}{:}{\U[m+1]}
  \eqnline{\Xp{m}}{}{\defeq}{\Xfrom{m}{0}[D=\unitpoint]}
  \graymidrule
  \eqnline{\Xfrom{m}{n}}{\eqnarg{X}{D}{\Xto{m}{n}}}{:}{\U[m+1]}
  \eqnline{\Xfrom{m}{n}}{D}{\defeq}{\Sigma R:\Xcomp{m}{n}[D=\D]. \Xfrom{m}{n+1}[D=\pair{D}{R}]}
\end{eqntable}

%% file: tab-core.tex
\providecommand{\lab}{tab:core}
\begin{eqntable}{Truncated $\nu$-sets, the core\label{\lab}}
  \eqnline{\Xto{m}{n}}{}{:}{\U[m+1]}
  \eqnline{\Xto{m}{0}}{}{\defeq}{\unittype}
  \eqnline{\Xto{m}{n'+1}}{}{\defeq}{\ensuremath{\Sigma}D:\Xto{m}{n'}.\,\Xcomp{m}{n'}[D=\D]}
  \graymidrule
  \eqnline{\Xcomp{m}{n}}{\eqnarg{X}{D}{\Xto{m}{n}}}{:}{\U[m+1]}
  \eqnline{\Xcomp{m}{n}}{D}{\defeq}{\fullframe[m][n][D=\D] \imp \U[m]}
\end{eqntable}

%% file: tab-frames.tex
\providecommand{\lab}{tab:frames}
\begin{eqntable}{$\mathsf{frame}$, $\mathsf{layer}$, and $\mathsf{painting}$\label{\lab}}
  \eqnline{\fullframe[m][n]}{\eqnarg{fullframe}{D}{\Xto{m}{n}}}{:}{\U[m]}
  \eqnline{\fullframe[m][n]}{D}{\defeq}{\framep[m][n][n][][D=\D]}
  \graymidrule

  \eqnline{\framep[m][n][p][p \leq n]}{\eqnarg{frame}{D}{\Xto{m}{n}}}{:}{\U[m]}
  \eqnline{\framep[m][n][0]}{D}{\defeq}{\unittype}
  \eqnline{\framep[m][n][p'+1]}{D}{\defeq}{\Sigma d:\framep[m][n][p'][][D=\D].\,\layer[m][n][p'][][D=\D, d=\d]}
  \graymidrule

  \eqnline{\layer[m][n][p][p < n]}{\makecell{\eqnarg{layer}{D}{\Xto{m}{n}} \\ \eqnarg{layer}{d}{\framep[m][n][p][][D=\D]}}}{:}{\U[m]}
  \eqnline{\layer[m][n][p]}{D~d}{\defeq}{\Pi \omega.\painting[m][n-1][p][][D=\hdD, E=\tlD, d={\restrf[m][\omega][p][n][p][][D=\D, d=\d]}]}
  \graymidrule

  \eqnline{\painting[m][n][p][p \leq n]}{\makecell{\eqnarg{filler}{D}{\Xto{m}{n}} \\ \eqnarg{filler}{E}{\Xcomp{m}{n}[D=\D]} \\ \eqnarg{filler}{d}{\framep[m][n][p][][D=\D]}}}{:}{\U[m]}
  \eqnline{\painting[m][n][p][p = n]}{\D~\E~\d}{\defeq}{\E(\d)}
  \eqnline{\painting[m][n][p][p < n]}{\D~\E~\d}{\defeq}{\Sigma l:\layer[m][n][p][][D=\D, d=\d].\,\painting[m][n][p+1][][D=\D, E=\E, d=\pair{\d}{\l}]}
\end{eqntable}

%% file: tab-faces-ett.tex
\providecommand{\lab}{fulltab:faces}
\begin{eqntable}{$q$-th projection of $\mathsf{restr}$, or faces\label{\lab}}
  \eqnline{\restrf[m][\epsilon][q][n][p][p \leq q \leq n - 1]}{\makecell{\eqnarg{restrframe}{D}{\Xto{m}{n}} \\ \eqnarg{restrframe}{d}{\framep[m][n][p][][D=\D]}}}{:}{\framep[m][n-1][p][][D=\hdD]}
  \eqnline{\restrf[m][\epsilon][q][n][0]}{D~\unitpoint}{\defeq}{\unitpoint}
  \eqnline{\restrf[m][\epsilon][q][n][p'+1]}{D~(\pair{d}{l})}{\defeq}{(\restrf[m][\epsilon][q][n][p'][][D=\D, d=\d],\restrl[m][\epsilon][q-1][n][p'][][D=\D, d=\d, l=\l])}
  \graymidrule

  \eqnline{\restrl[m][\epsilon][q][n][p][p \leq q \leq n - 2]}{\makecell{\eqnarg{restrlayer}{D}{\Xto{m}{n}} \\ \eqnarg{restrlayer}{d}{\framep[m][n][p][][D=\D]} \\ \eqnarg{restrlayer}{l}{\layer[m][n][p][][D=\D, d=\d]}}}{:}{\layer[m][n-1][p][][D=\hdD, d={\restrf[m][\epsilon][q+1][n][p][][D=\D, d=\d]}]}
  \eqnline{\restrl[m][\epsilon][q][n,p]}{\D~\d~\l}{\defeq}{\lambda \omega.(\restrc[m][\epsilon][q][n-1][p][][D=\hdD, E=\tlD, d={\restrf[m][\omega][p][n][p][][D=\D, d=\d]}, c={\l_\omega}])}
  \graymidrule

  \eqnline{\restrc[m][\epsilon][q][n][p][p \leq q \leq n - 1]}{\makecell{\eqnarg{restrfiller}{D}{\Xto{m}{n}} \\ \eqnarg{restrfiller}{E}{\Xcomp{m}{n}[D=\D]} \\\eqnarg{restrfiller}{d}{\framep[m][n][p][][D=\D]} \\ \eqnarg{restrfiller}{c}{\painting[m][n][p][][D=\D, E=\E, d=\d]}}}{:}{\painting[m][n-1][p][][D=\hdD, E=\tlD, d={\restrf[m][\epsilon][q][n][p][][D=\D, d=\d]}]}
  \eqnline{\restrc[m][\epsilon][q][n][p][p=q]}{\D~\E~\d~(\pair{l}{\_})}{\defeq}{\l_\epsilon}
  \eqnline{\restrc[m][\epsilon][q][n][p][p<q]}{\D~\E~\d~(\pair{l}{c})}{\defeq}{(\restrl[m][\epsilon][q-1][n][p][][D=\D, d=\d, l=\l],\restrc[m][\epsilon][q][n][p+1][][D=\D, E=\E, d=\pair{d}{l}, c=\c])}
\end{eqntable}

%% file: tab-coh-ett.tex
\providecommand{\lab}{fulltab:coh}
\begin{eqntable}{Commutation of $q$-th projection and $r$-th projection, or coherence conditions\label{\lab}}
  \eqnline{\cohf[m][\epsilon][\omega][q][r][n][p][p \leq r \leq q \leq n - 2]}{\makecell{\eqnarg{cohframe}{D}{\Xto{m}{n}} \\ \eqnarg{cohframe}{d}{\framep[m][n][p][][D=\D]}}}{:}{\makecell{\restrf[m][\epsilon][q][n-1][p][][D=\hdD, d={\restrf[m][\omega][r][n][p][][D=\D, d=\d]}] \\ = \restrf[m][\omega][r][n-1][p][][D=\hdD, d={\restrf[m][\epsilon][q+1][n][p][][D=\D, d=\d]}]}}
  \eqnline{\cohf[m][\epsilon][\omega][q][r][n][0]}{\D~\unitpoint}{\defeq}{\refl(\unitpoint)}
  \eqnline{\cohf[m][\epsilon][\omega][q][r][n][p'+1]}{\D~(\pair{\d}{\l})}{\defeq}{(\cohf[m][\epsilon][\omega][q][r][n][p'][][D=\D, d=\d], \cohl[m][\epsilon][\omega][q-1][r-1][n][p'][][D=\D, d=\d, l=\l])}
  \graymidrule

  \eqnline{\cohl[m][\epsilon][\omega][q][r][n][p][p \leq r \leq q \leq n - 3]}{\makecell{\eqnarg{cohlayer}{D}{\Xto{m}{n}} \\ \eqnarg{cohlayer}{d}{\framep[m][n][p][][D=\D]} \\ \eqnarg{cohlayer}{l}{\layer[m][n][p][][D=\D, d=\d]}}}{:}{\makecell{\restrl[m][\epsilon][q][n-1][p][][D=\hdD, d={\restrf[m][\omega][r][n][p][][D=\D, d=\d]}](\restrl[m][\omega][r][n][p][][D=\D, d=\d, l=\l]) \\ = \restrl[m][\omega][r][n-1][p][][D=\hdD, d={\restrf[m][\epsilon][q+1][n][p][][D=\D, d=\d]}, l={\restrl[m][\epsilon][q+1][n][p][][D=\D, d=\d, l=\l]}]}}
  \eqnline{\cohl[m][\epsilon][\omega][q][r][n][p]}{\D~\d~\l}{\defeq}{\makecell{\lambda \theta .\;\cohc[m][\epsilon][\omega][q][r][n-1][p][][E=\tlD, c=\l_\theta]}}
  \graymidrule

  \eqnline{\cohc[m][\epsilon][\omega][q][r][n][p][p \leq r \leq q \leq n - 2]}{\makecell{\eqnarg{cohpainting}{D}{\Xto{m}{n}} \\ \eqnarg{cohpainting}{E}{\Xcomp{m}{n}[D=\D]} \\ \eqnarg{cohpainting}{d}{\framep[m][n][p][][D=\D]} \\ \eqnarg{cohpainting}{c}{\painting[m][n][p][][D=\D, E=\E, d=\d]}}}{:}{\makecell{\restrc[m][\epsilon][q][n-1][p][][D=\hdD, E=\tlD, d={\restrf[m][\omega][r][n][p][][D=\D, d=\d]}, c={\restrc[m][\omega][r][n][p][][D=\D, E=\E, d=\d, c=\c]}] \\ = \restrc[m][\omega][r][n-1][p][][D=\hdD, E=\tlD, D={\restrf[m][\epsilon][q+1][n][p][][D=\D, d=\d]}, c={\restrc[m][\epsilon][q+1][n][p][][D=\D, E=\E, d=\d, c=\c]}]}}
  \eqnline{\cohc[m][\epsilon][\omega][q][r][n][p][p=r]}{\D~\E~\d~(\pair{\l}{\_})}{\defeq}{\refl(\restrc[m][\epsilon][q-1][n-1][p][][D=\hdD, E=\tlD, d={\restrf[m][\omega][p][n][p][][D=\D, d=\d]}, c={\l_\epsilon}])}
  \eqnline{\cohc[m][\epsilon][\omega][q][r][n][p][p<r]}{\D~\E~\d~(\pair{\l}{\c})}{\defeq}{(\pair{\cohl[m][\epsilon][\omega][q][r][n][p][][D=\D, d=\d, l=\l]}{\cohc[m][\epsilon][\omega][q][r][n][p+1][][D=\D, E=\E, d=\pair{d}{l}, c=\c]})}
\end{eqntable}

%% file: tab-faces.tex
\providecommand{\lab}{tab:faces}
\begin{eqntable}{$q$-th projection of $\mathsf{restr}$, or faces\label{\lab}}
  \eqnline{\restrf[m][\epsilon][q][n][p][p \leq q \leq n - 1]}{\makecell{\eqnarg{restrframe}{D}{\Xto{m}{n}}  \\ \eqnarg{restrframe}{d}{\framep[m][n][p][][D=\D]}}}{:}{\framep[m][n-1][p][][D=\hdD]}
  \eqnline{\restrf[m][\epsilon][q][n][0]}{D~\unitpoint}{\defeq}{\unitpoint}
  \eqnline{\restrf[m][\epsilon][q][n][p'+1]}{D~(\pair{d}{l})}{\defeq}{(\restrf[m][\epsilon][q][n][p'][][D=\D, d=\d],\restrl[m][\epsilon][q-1][n][p'][][D=\D, d=\d, l=\l])}
  \graymidrule

  \eqnline{\restrl[m][\epsilon][q][n][p][p \leq q \leq n - 2]}{\makecell{\eqnarg{restrlayer}{D}{\Xto{m}{n}}  \\ \eqnarg{restrlayer}{d}{\framep[m][n][p][][D=\D]} \\ \eqnarg{restrlayer}{l}{\layer[m][n][p][][D=\D, d=\d]}}}{:}{\layer[m][n-1][p][][D=\hdD, d={\restrf[m][\epsilon][q+1][n][p][][D=\D, d=\d]}]}
  \eqnline{\restrl[m][\epsilon][q][n,p]}{\D~\d~\l}{\defeq}{\lambda \omega.(\overright{\cohf[m][\epsilon][\omega][q][p][n][p][][D=\D, d=\d]}(\restrc[m][\epsilon][q][n-1][p][][D=\hdD, E=\tlD, d={\restrf[m][\omega][p][n][p][][D=\D, d=\d]}, c={\l_\omega}]))}
  \graymidrule

  \eqnline{\restrc[m][\epsilon][q][n][p][p \leq q \leq n - 1]}{\makecell{\eqnarg{restrfiller}{D}{\Xto{m}{n}} \\ \eqnarg{restrfiller}{E}{\Xcomp{m}{n}[D=\D]} \\\eqnarg{restrfiller}{d}{\framep[m][n][p][][D=\D]} \\ \eqnarg{restrfiller}{c}{\painting[m][n][p][][D=\D, E=\E, d=\d]}}}{:}{\painting[m][n-1][p][][D=\hdD, E=\tlD, d={\restrf[m][\epsilon][q][n][p][][D=\D, d=\d]}]}
  \eqnline{\restrc[m][\epsilon][q][n][p][p=q]}{\D~\E~\d~(\pair{l}{\_})}{\defeq}{\l_\epsilon}
  \eqnline{\restrc[m][\epsilon][q][n][p][p<q]}{\D~\E~\d~(\pair{l}{c})}{\defeq}{(\restrl[m][\epsilon][q-1][n][p][][D=\D, d=\d, l=\l],\restrc[m][\epsilon][q][n][p+1][][D=\D, E=\E, d=\pair{d}{l}, c=\c])}
\end{eqntable}

%% file: tab-coh.tex
\providecommand{\lab}{tab:coh}
\begin{eqntable}{Commutation of $q$-th projection and $r$-th projection, or coherence conditions\label{\lab}}
  \eqnline{\cohf[m][\epsilon][\omega][q][r][n][p][p \leq r \leq q \leq n - 2]}{\makecell{\eqnarg{cohframe}{D}{\Xto{m}{n}}                                                                                                                                                                                                  \\ \eqnarg{cohframe}{d}{\framep[m][n][p][][D=\D]}}}{:}{\makecell{\restrf[m][\epsilon][q][n-1][p][][D=\hdD, d={\restrf[m][\omega][r][n][p][][D=\D, d=\d]}] \\ = \restrf[m][\omega][r][n-1][p][][D=\hdD, d={\restrf[m][\epsilon][q+1][n][p][][D=\D, d=\d]}]}}
  \eqnline{\cohf[m][\epsilon][\omega][q][r][n][0]}{\D~\unitpoint}{\defeq}{\refl(\unitpoint)}
  \eqnline{\cohf[m][\epsilon][\omega][q][r][n][p'+1]}{\D~(\pair{\d}{\l})}{\defeq}{(\cohf[m][\epsilon][\omega][q][r][n][p'][][D=\D, d=\d], \cohl[m][\epsilon][\omega][q-1][r-1][n][p'][][D=\D, d=\d, l=\l])}
  \graymidrule

  \eqnline{\cohl[m][\epsilon][\omega][q][r][n][p][p \leq r \leq q \leq n - 3]}{\makecell{\eqnarg{cohlayer}{D}{\Xto{m}{n}}                                                                                                                                                                                                                       \\ \eqnarg{cohlayer}{d}{\framep[m][n][p][][D=\D]} \\ \eqnarg{cohlayer}{l}{\layer[m][n][p][][D=\D, d=\d]}}}{:}{\makecell{\overright{\cohf[m][\epsilon][\omega][q+1][r+1][n][p][][D=\D, d=\d]}(\restrl[m][\epsilon][q][n-1][p][][D=\hdD, d={\restrf[m][\omega][r][n][p][][D=\D, d=\d]}](\restrl[m][\omega][r][n][p][][D=\D, d=\d, l=\l])) \\ = \restrl[m][\omega][r][n-1][p][][D=\hdD, d={\restrf[m][\epsilon][q+1][n][p][][D=\D, d=\d]}, l={\restrl[m][\epsilon][q+1][n][p][][D=\D, d=\d, l=\l]}]}}
  \eqnline{\cohl[m][\epsilon][\omega][q][r][n][p]}{\D~\d~\l}{\defeq}{\makecell{\lambda \theta . (\overright{\coht[m][\epsilon][\omega][\theta][q][r][n][p][d=\d]})(\ap(\overright{\cohf[m][\omega][\theta][r][p][n-1][p][][D=\hdD, E=\tlD, d={\restrf[m][\epsilon][q+2][n][p][][D=\D, d=\d, l=\l_\theta]}, c=\l_\theta]})) \\ (\ap(\overright{\restrf[m][\omega][r][n-2][p][][D=\D, d={\cohf[m][\epsilon][\theta][q+1][p][n][p][]}]}))\;\cohc[m][\epsilon][\omega][q][r][n-1][p][][E=\tlD, c=\l_\theta]}}
  \graymidrule

  \eqnline{\cohc[m][\epsilon][\omega][q][r][n][p][p \leq r \leq q \leq n - 2]}{\makecell{\eqnarg{cohpainting}{D}{\Xto{m}{n}}                                                                                                                                                                                                 \\ \eqnarg{cohpainting}{E}{\Xcomp{m}{n}[D=\D]} \\ \eqnarg{cohpainting}{d}{\framep[m][n][p][][D=\D]} \\ \eqnarg{cohpainting}{c}{\painting[m][n][p][][D=\D, E=\E, d=\d]}}}{:}{\makecell{\overright{\cohf[m][\epsilon][\omega][q][r][n][p][][D=\D, d=\d]}(\restrc[m][\epsilon][q][n-1][p][][D=\hdD, E=\tlD, d={\restrf[m][\omega][r][n][p][][D=\D, d=\d]}, c={\restrc[m][\omega][r][n][p][][D=\D, E=\E, d=\d, c=\c]}]) \\ = \restrc[m][\omega][r][n-1][p][][D=\hdD, E=\tlD, D={\restrf[m][\epsilon][q+1][n][p][][D=\D, d=\d]}, c={\restrc[m][\epsilon][q+1][n][p][][D=\D, E=\E, d=\d, c=\c]}]}}
  \eqnline{\cohc[m][\epsilon][\omega][q][r][n][p][p=r]}{\D~\E~\d~(\pair{\l}{\_})}{\defeq}{\refl(\restrc[m][\epsilon][q-1][n-1][p][][D=\hdD, E=\tlD, d={\restrf[m][\omega][p][n][p][][D=\D, d=\d]}, c={\l_\epsilon}])}
  \eqnline{\cohc[m][\epsilon][\omega][q][r][n][p][p<r]}{\D~\E~\d~(\pair{\l}{\c})}{\defeq}{(\pair{\cohl[m][\epsilon][\omega][q][r][n][p][][D=\D, d=\d, l=\l]}{\cohc[m][\epsilon][\omega][q][r][n][p+1][][D=\D, E=\E, d=\pair{d}{l}, c=\c]})}
\end{eqntable}

%% file: paper.bbl
\newcommand{\etalchar}[1]{$^{#1}$}
\begin{thebibliography}{CCHM18}

\bibitem[ABC{\etalchar{+}}21]{angiuli21}
Carlo Angiuli, Guillaume Brunerie, Thierry Coquand, Robert Harper, Kuen-Bang
  Hou~(Favonia), and Daniel~R. Licata.
\newblock Syntax and models of cartesian cubical type theory.
\newblock {\em Math. Struct. Comput. Sci.}, 31(4):424--468, 2021.

\bibitem[ACK16]{altenkirch16}
Thorsten Altenkirch, Paolo Capriotti, and Nicolai Kraus.
\newblock Extending homotopy type theory with strict equality.
\newblock In Jean{-}Marc Talbot and Laurent Regnier, editors, {\em 25th {EACSL}
  Annual Conference on Computer Science Logic, {CSL} 2016, August 29 -
  September 1, 2016, Marseille, France}, volume~62 of {\em LIPIcs}, pages
  21:1--21:17. Schloss Dagstuhl - Leibniz-Zentrum f{\"{u}}r Informatik, 2016.

\bibitem[ACK17]{annenkovCK17}
Danil Annenkov, Paolo Capriotti, and Nicolai Kraus.
\newblock Two-level type theory and applications.
\newblock {\em CoRR}, abs/1705.03307, 2017.

\bibitem[ACKS23]{AnnenkovCKS2023}
Danil Annenkov, Paolo Capriotti, Nicolai Kraus, and Christian Sattler.
\newblock Two-level type theory and applications.
\newblock {\em Mathematical Structures in Computer Science}, 33(8):688–743,
  2023.

\bibitem[AGJ14]{atkey14}
Robert Atkey, Neil Ghani, and Patricia Johann.
\newblock A relationally parametric model of dependent type theory.
\newblock In {\em The 41st Annual ACM SIGPLAN-SIGACT Symposium on Principles of
  Programming Languages, POPL '14, San Diego, CA, USA, January 20-21, 2014.},
  2014.

\bibitem[AK15]{altenkirch15}
Thorsten Altenkirch and Ambrus Kaposi.
\newblock Towards a cubical type theory without an interval.
\newblock In Tarmo Uustalu, editor, {\em 21st International Conference on Types
  for Proofs and Programs, {TYPES} 2015, May 18-21, 2015, Tallinn, Estonia},
  volume~69 of {\em LIPIcs}, pages 3:1--3:27. Schloss Dagstuhl -
  Leibniz-Zentrum f{\"{u}}r Informatik, 2015.

\bibitem[BCH13a]{bezem13}
Marc Bezem, Thierry Coquand, and Simon Huber.
\newblock A model of type theory in cubical sets.
\newblock In Ralph Matthes and Aleksy Schubert, editors, {\em 19th
  International Conference on Types for Proofs and Programs, {TYPES} 2013,
  April 22-26, 2013, Toulouse, France}, volume~26 of {\em LIPIcs}, pages
  107--128. Schloss Dagstuhl - Leibniz-Zentrum fuer Informatik, 2013.

\bibitem[BCH13b]{BezemCoquandHuber13}
Marc Bezem, Thierry Coquand, and Simon Huber.
\newblock A model of type theory in cubical sets.
\newblock In Ralph Matthes and Aleksy Schubert, editors, {\em 19th
  International Conference on Types for Proofs and Programs, {TYPES} 2013,
  April 22-26, 2013, Toulouse, France}, volume~26 of {\em LIPIcs}, pages
  107--128. Schloss Dagstuhl - Leibniz-Zentrum fuer Informatik, 2013.

\bibitem[BCM15]{bernardy15}
Jean{-}Philippe Bernardy, Thierry Coquand, and Guilhem Moulin.
\newblock A presheaf model of parametric type theory.
\newblock {\em Electr. Notes Theor. Comput. Sci.}, 319:67--82, 2015.

\bibitem[BJP10]{bernardy10}
Jean{-}Philippe Bernardy, Patrik Jansson, and Ross Paterson.
\newblock Parametricity and dependent types.
\newblock In Paul Hudak and Stephanie Weirich, editors, {\em Proceeding of the
  15th {ACM} {SIGPLAN} international conference on Functional programming,
  {ICFP} 2010, Baltimore, Maryland, USA, September 27-29, 2010}, pages
  345--356. {ACM}, 2010.

\bibitem[BL11]{bernardy11}
Jean{-}Philippe Bernardy and Marc Lasson.
\newblock Realizability and parametricity in pure type systems.
\newblock In {\em Foundations of Software Science and Computational Structures
  - 14th International Conference, {FOSSACS} 2011, Held as Part of the Joint
  European Conferences on Theory and Practice of Software, {ETAPS} 2011,
  Saarbr{\"{u}}cken, Germany, March 26-April 3, 2011. Proceedings}, volume 6604
  of {\em Lecture Notes in Computer Science}, pages 108--122. Springer, 2011.

\bibitem[BM12]{bernardy12}
Jean-Philippe Bernardy and Guilhem Moulin.
\newblock A computational interpretation of parametricity.
\newblock In {\em 2012 27th Annual IEEE Symposium on Logic in Computer
  Science}, pages 135--144. IEEE, 2012.

\bibitem[BM17]{buchholtz17}
Ulrik Buchholtz and Edward Morehouse.
\newblock Varieties of cubical sets.
\newblock In {\em International Conference on Relational and Algebraic Methods
  in Computer Science}, pages 77--92. Springer, 2017.

\bibitem[CCHM15]{cchm}
Cyril Cohen, Thierry Coquand, Simon Huber, and Anders M{\"{o}}rtberg.
\newblock Cubical type theory: {A} constructive interpretation of the
  univalence axiom.
\newblock In Tarmo Uustalu, editor, {\em 21st International Conference on Types
  for Proofs and Programs, {TYPES} 2015, May 18-21, 2015, Tallinn, Estonia},
  volume~69 of {\em LIPIcs}, pages 5:1--5:34. Schloss Dagstuhl -
  Leibniz-Zentrum f{\"{u}}r Informatik, 2015.

\bibitem[CCHM18]{cohen16}
Cyril Cohen, Thierry Coquand, Simon Huber, and Anders M{\"{o}}rtberg.
\newblock {Cubical Type Theory: A Constructive Interpretation of the Univalence
  Axiom}.
\newblock In Tarmo Uustalu, editor, {\em TYPES 2015}, volume~69 of {\em
  LIPIcs}, pages 5:1--5:34. Schloss Dagstuhl, 2018.

\bibitem[CGH14]{curien14}
Pierre-Louis Curien, Richard Garner, and Martin Hofmann.
\newblock Revisiting the categorical interpretation of dependent type theory.
\newblock {\em Theoretical Computer Science}, 546:99--119, 2014.
\newblock Models of Interaction: Essays in Honour of Glynn Winskel.

\bibitem[CH20]{cavallo19}
Evan Cavallo and Robert Harper.
\newblock {Internal Parametricity for Cubical Type Theory}.
\newblock In Maribel Fern{\'a}ndez and Anca Muscholl, editors, {\em 28th EACSL
  Annual Conference on Computer Science Logic (CSL 2020)}, volume 152 of {\em
  Leibniz International Proceedings in Informatics (LIPIcs)}, pages
  13:1--13:17, Dagstuhl, Germany, 2020. Schloss Dagstuhl--Leibniz-Zentrum fuer
  Informatik.

\bibitem[GCST19]{gilbert19}
Ga{\"{e}}tan Gilbert, Jesper Cockx, Matthieu Sozeau, and Nicolas Tabareau.
\newblock Definitional proof-irrelevance without {K}.
\newblock {\em Proc. {ACM} Program. Lang.}, 3({POPL}):3:1--3:28, 2019.

\bibitem[GM03]{grandis03}
Marco Grandis and Luca Mauri.
\newblock Cubical sets and their site.
\newblock {\em Theory and Applications of Categories}, 11:185--211, 2003.

\bibitem[Her15]{herbelin15}
Hugo Herbelin.
\newblock A dependently-typed construction of semi-simplicial types.
\newblock {\em Mathematical Structures in Computer Science}, 25(5):1116--1131,
  2015.

\bibitem[Her20]{herbelin-hott-uf}
Hugo Herbelin.
\newblock Investigations into syntactic iterated parametricity and cubical type
  theory, 2020.
\newblock Available at
  \url{https://hott-uf.github.io/2020/HoTTUF_2020_paper_22.pdf}.

\bibitem[Hof95]{HofmannPhd}
Martin Hofmann.
\newblock {\em Extensional concepts in intensional type theory}.
\newblock PhD thesis, University of Edinburgh, 1995.

\bibitem[HS94]{HofmannStreicher94}
Martin Hofmann and Thomas Streicher.
\newblock The groupoid model refutes uniqueness of identity proofs.
\newblock In {\em Proceedings of the Ninth Annual Symposium on Logic in
  Computer Science {(LICS} '94), Paris, France, July 4-7, 1994}, pages
  208--212. {IEEE} Computer Society, 1994.

\bibitem[JS17]{johann17}
Patricia Johann and Kristina Sojakova.
\newblock Cubical categories for higher-dimensional parametricity.
\newblock {\em CoRR}, abs/1701.06244, 2017.

\bibitem[KL21]{kapulkin21}
Krzysztof Kapulkin and Peter~LeFanu Lumsdaine.
\newblock The simplicial model of {Univalent Foundations} (after {Voevodsky}).
\newblock {\em Journal of the European Mathematical Society}, 23(6):2071--2126,
  2021.

\bibitem[Kra21]{kraus21}
Nicolai Kraus.
\newblock Internal {\(\infty\)}-categorical models of dependent type theory :
  Towards {2LTT} eating {HoTT}.
\newblock In {\em 36th Annual {ACM/IEEE} Symposium on Logic in Computer
  Science, {LICS} 2021, Rome, Italy, June 29 - July 2, 2021}, pages 1--14.
  {IEEE}, 2021.

\bibitem[KS17]{kraus17}
Nicolai Kraus and Christian Sattler.
\newblock Space-valued diagrams, type-theoretically.
\newblock {\em arXiv preprint arXiv:1704.04543}, 2017.

\bibitem[Las14]{lasson12}
Marc Lasson.
\newblock {\em R\'ealisabilit\'e et Param\'etricit\'e dans les Syst\`emes de
  Types Purs}.
\newblock PhD thesis, Universit\'e de Lyon, Nov 2014.

\bibitem[LR20]{LoregianRiehl20}
Fosco Loregian and Emily Riehl.
\newblock Categorical notions of fibration.
\newblock {\em Expositiones Mathematicae}, 38(4):496--514, 2020.

\bibitem[Mar84]{martinlof84}
Per Martin{-}L{\"{o}}f.
\newblock {\em Intuitionistic type theory}, volume~1 of {\em Studies in proof
  theory}.
\newblock Bibliopolis, 1984.

\bibitem[ML75]{martinlof75}
Per Martin-L\"of.
\newblock An intuitionistic theory of types, predicative part.
\newblock In {\em Logic Colloquium}, pages 73--118. North Holland, 1975.

\bibitem[Moe21]{moeneclaey21}
Hugo Moeneclaey.
\newblock Parametricity and semi-cubical types.
\newblock In {\em 36th Annual {ACM/IEEE} Symposium on Logic in Computer
  Science, {LICS} 2021, Rome, Italy, June 29 - July 2, 2021}, pages 1--11.
  {IEEE}, 2021.

\bibitem[Moe22]{moeneclaey22phd}
Hugo Moeneclaey.
\newblock {\em Cubical models are cofreely parametric}.
\newblock PhD thesis, Universit\'e Paris Cit\'e, 2022.

\bibitem[Mou16]{moulin16}
Guilhem Moulin.
\newblock {\em Internalizing parametricity}.
\newblock PhD thesis, Department of Computer Science and Engineering, Chalmers
  University of Technology, 2016.

\bibitem[ND24]{nuytsdevriese24}
Andreas Nuyts and Dominique Devriese.
\newblock {Transpension: The Right Adjoint to the Pi-type}.
\newblock {\em {Logical Methods in Computer Science}}, {Volume 20, Issue 2},
  June 2024.

\bibitem[NVD17]{nuyts17}
Andreas Nuyts, Andrea Vezzosi, and Dominique Devriese.
\newblock Parametric quantifiers for dependent type theory.
\newblock {\em Proc. ACM Program. Lang.}, 1(ICFP), aug 2017.

\bibitem[PL15]{part15}
Fedor Part and Zhaohui Luo.
\newblock Semi-simplicial types in logic-enriched homotopy type theory.
\newblock {\em CoRR}, abs/1506.04998, 2015.

\bibitem[Rey83]{reynolds83}
John~C. Reynolds.
\newblock Types, abstraction and parametric polymorphism.
\newblock In R.~E.~A. Mason, editor, {\em Information Processing 83,
  Proceedings of the {IFIP} 9th World Computer Congress, Paris, France,
  September 19-23, 1983}, pages 513--523. North-Holland/IFIP, 1983.

\bibitem[Shu15]{shulman15}
Michael Shulman.
\newblock Univalence for inverse diagrams and homotopy canonicity.
\newblock {\em Mathematical Structures in Computer Science}, 25(5):1203–1277,
  2015.

\bibitem[{The}13]{hottbook}
{The Univalent Foundations Program}.
\newblock {\em Homotopy Type Theory: Univalent Foundations of Mathematics}.
\newblock Institute for Advanced Study, 2013.

\bibitem[Voe12]{voevodsky12}
Vladimir Voevodsky.
\newblock Semi-simplicial types, Nov 2012.
\newblock Briefly described in Section~8 of \cite{herbelin15}.

\end{thebibliography}
